\documentclass[a4paper,11pt]{article}
\pdfoutput=1 

\usepackage{jheppub} 

\usepackage[T1]{fontenc} 

\usepackage{graphicx}
\usepackage{xspace}
\usepackage{amsmath}
\usepackage{amsfonts}
\usepackage{amssymb}
\usepackage{enumerate}
\usepackage{epsfig}
\usepackage{lineno}
\usepackage{xcolor}

\newcommand{\ksn}{\ensuremath{\mathrm{K_S}}\xspace}

\newcommand{\ksnkln}{\ensuremath{\mathrm{K_S K_L}}\xspace}
\newcommand{\ksln}{\ensuremath{\mathrm{K_{S,L}}}\xspace}
\newcommand{\klsn}{\ensuremath{\mathrm{K_{L,S}}}\xspace}
\newcommand{\kln}{\ensuremath{\mathrm{K_L}}\xspace}

\newcommand{\kn}{\ensuremath{\mathrm{K^0}}\xspace}
\newcommand{\knped}{\ensuremath{\mathrm{K}}\xspace}
\newcommand{\knknb}{\ensuremath{\mathrm{K^0\bar{K}^0}}\xspace}
\newcommand{\knb}{\ensuremath{\mathrm{\bar{K}^0}}\xspace}

\def\CP                {\ensuremath{\mathcal{CP}}\xspace}
\def\CPT               {\ensuremath{\mathcal{CPT}}\xspace} 
\def\C       {\ensuremath{\mathcal{C}}\xspace}
\def\P       {\ensuremath{\mathcal{P}}\xspace}
\def\T       {\ensuremath{\mathcal{T}}\xspace}

\newcommand{\pvec}{\ensuremath{ \vec{p} }}

\newcommand{\sqrts}{\ensuremath{\sqrt{s}}}

\newcommand{\Pphi}{\ensuremath{\phi}}

\def\ifm#1{\relax\ifmmode#1\else$#1$\fi}  
\def\x{\ifm{\times}}  \def\pt#1,#2,{\ifm{#1\x10^{#2}}}

\def\to{\ifm{\rightarrow}}

\def\rmk{\rm\kern.5mm }   \def\f{\ifm{\phi}}  

\title{\boldmath 
Precision tests of Quantum
Mechanics
and \CPT symmetry
with entangled neutral kaons
 at KLOE
}
\collaboration{The KLOE-2 Collaboration}
\newcommand{\affuni}[2]{Dipartimento di Fisica dell'Universit\`a #1, #2, Italy.}
\newcommand{\affinfn}[2]{INFN Sezione di #1, #2, Italy.}

\author[d]{D.~Babusci}
\author[v]{M.~Berlowski}
\author[d]{C.~Bloise}
\author[d]{F.~Bossi}
\author[s]{P.~Branchini}
\author[r,s]{A.~Budano}
\author[u]{B.~Cao}
\author[r,s]{F.~Ceradini}
\author[d]{P.~Ciambrone}
\author[j,k]{F.~Curciarello}
\author[c]{E.~Czerwi\'nski}
\author[n,o]{G.~D'Agostini}
\author[n,o]{R.~D'Amico}
\author[d]{E.~Dan\`e}
\author[n,o]{V.~De~Leo}
\author[d]{E.~De~Lucia}
\author[d]{A.~De~Santis}
\author[d]{P.~De~Simone}
\author[r,s]{A.~Di~Cicco}
\author[n,o]{A.~Di~Domenico}
\emailAdd{antonio.didomenico@roma1.infn.it}
\author[d]{E.~Diociaiuti}
\author[d]{D.~Domenici}
\author[d]{A.~D'Uffizi}
\author[p,q]{A.~Fantini}
\author[n,o]{G.~Fantini}
\author[d]{P.~Fermani}
\author[t,o]{S.~Fiore}
\author[c]{A.~Gajos}
\author[n,o]{P.~Gauzzi}
\author[d]{S.~Giovannella}
\author[s]{E.~Graziani}
\author[g,h]{V.~L.~Ivanov}
\author[u]{T.~Johansson}
\author[d, w]{X.~Kang}
\author[c]{D.~Kisielewska-Kami\'nska}
\author[g,h]{E.~A.~Kozyrev}
\author[v]{W.~Krzemien}
\author[u]{A.~Kupsc}
\author[g,h]{P.~A.~Lukin}
\author[f,b]{G.~Mandaglio}
\author[d,m]{M.~Martini}
\author[p,q]{R.~Messi}
\author[d]{S.~Miscetti}
\author[d]{D.~Moricciani}
\author[c]{P.~Moskal}
\author[s]{A.~Passeri}
\author[l,o]{V.~Patera}
\author[n,o]{E.~Perez~del~Rio}
\author[d]{P.~Santangelo}
\author[j,k]{M.~Schioppa}
\author[r,s]{A.~Selce}
\author[c]{M.~Silarski}
\author[d,e]{F.~Sirghi}
\author[g,h]{E.~P.~Solodov}
\author[s]{L.~Tortora}
\author[i]{G.~Venanzoni}
\author[v]{W.~Wi\'slicki}
\author[u]{M.~Wolke}

\affiliation[a]{Dipartimento di Fisica e Astronomia ``Ettore Majorana'',
Universit\`a di Catania, Italy}
\affiliation[b]{\affinfn{Catania}{Catania}}
\affiliation[c]{Institute of Physics, Jagiellonian University, Cracow, Poland.}
\affiliation[d]{Laboratori Nazionali di Frascati dell'INFN, Frascati, Italy.}
\affiliation[e]{Horia Hulubei National Institute of Physics and Nuclear Engineering, M\v{a}gurele, Romania.}
\affiliation[f]{Dipartimento di Scienze Matematiche e Informatiche, Scienze Fisiche e Scienze della Terra dell'Universit\`a di Messina, Messina, Italy.}
\affiliation[g]{Budker Institute of Nuclear Physics, Novosibirsk, Russia.}
\affiliation[h]{Novosibirsk State University, Novosibirsk, Russia.}
\affiliation[i]{\affinfn{Pisa}{Pisa}}
\affiliation[j]{\affuni{della Calabria}{Arcavacata di Rende}}
\affiliation[k]{INFN Gruppo collegato di Cosenza, Arcavacata di Rende, Italy.}
\affiliation[l]{Dipartimento di Scienze di Base ed Applicate per l'Ingegneria dell'Universit\`a ``Sapienza'', Roma, Italy.}
\affiliation[m]{Dipartimento di Scienze e Tecnologie applicate, Universit\`a ``Guglielmo Marconi'', Roma, Italy.}
\affiliation[n]{\affuni{``Sapienza''}{Roma}}
\affiliation[o]{\affinfn{Roma}{Roma}}
\affiliation[p]{\affuni{``Tor Vergata''}{Roma}}
\affiliation[q]{\affinfn{Roma Tor Vergata}{Roma}}
\affiliation[r]{Dipartimento di Matematica e Fisica dell'Universit\`a 
``Roma Tre'', Roma, Italy.}
\affiliation[s]{\affinfn{Roma Tre}{Roma}}
\affiliation[t]{ENEA, Department of Fusion and Technology for Nuclear Safety and Security, Frascati (RM), Italy.}
\affiliation[u]{Department of Physics and Astronomy, Uppsala University, Uppsala, Sweden.}
\affiliation[v]{National Centre for Nuclear Research, Warsaw, Poland.}
\affiliation[w]{School of Mathematics and Physics, China University of Geosciences (Wuhan), Wuhan, China.}

\abstract{
The 
quantum interference 
between the decays 
of entangled neutral kaons is studied
in the process $\phi\rightarrow\ksn\kln\rightarrow\pi^+\pi^-\pi^+\pi^-$, which
exhibits 
the characteristic Einstein--Podolsky--Rosen correlations that
prevent
both kaons to decay into $\pi^+\pi^-$ at the same time.
This constitutes
a 
very 
powerful
tool for 
testing at the utmost precision
the quantum coherence of the entangled 
kaon pair state,
and to search for tiny
decoherence and \CPT violation effects,
which 
may be 
justified 
in a quantum gravity framework.
\par
The analysed data sample 
was
collected with the KLOE detector 
at 
DA$\Phi$NE, 
the Frascati $\phi$-factory,
and corresponds to an integrated
luminosity of about 1.7 fb$^{-1}$, i.e. 
to 
{about
$1.7 \times 10^9$
$\phi\rightarrow\ksnkln$
decays produced.}
From the fit of the observed $\Delta t$ distribution, 
being
$\Delta t$
the difference of the kaon decay times, 
the decoherence and \CPT violation parameters of various phenomenological models are measured with a largely improved
accuracy with respect to previous analyses.
\par
The results are consistent with no deviation from quantum mechanics 
and \CPT symmetry, 
while for some parameters the precision 
reaches
the interesting level at which -- in the most optimistic scenarios -- quantum gravity effects might show up.
They
provide
the most stringent limits up to date on the 
considered
models. 
}

\begin{document} 
\maketitle
\flushbottom

\section {Introduction}
\label{sec:intro}
Entanglement is one of the most striking features of quantum mechanics. 
Schr\"odinger, in reply to the famous argument by 
Einstein, 
Podolsky and 
Rosen (EPR) \cite{epr}, named {\it entanglement} the (non-local) correlations among the parts of a composite quantum system, and 
wrote:
``I would call this not {\it one} but rather {\it the} characteristic trait of quantum mechanics, the one that enforces its entire departure from classical lines of thought” \cite{schro}.
%
\par 
Correlations in \knknb pairs 
produced in
 a $\C=-1$ state,
 {where \C represents charge conjugation,}
were firstly recognised to be of the EPR-type 
by 
Lee and 
Yang in 1960~\cite{leeyang,day,inglis}. 
The entanglement of $\kn\knb$ pairs produced in $\phi$-meson decays, 
in combination with the unique properties of the neutral kaon system --
{such} 
as flavour oscillations, charge-parity (\CP) and
 time-reversal  (\T) violation --
%
%
%
%
opens up new horizons
in testing the basic principles of quantum mechanics and its 
fundamental discrete symmetries \T, \CP, \CPT~\cite{lipkin68,dunietz,buchanan,isidori,didohand,ktviol,kcptviol}.
%
%
%
%
%
%
%
\par
Neutral kaon pairs are produced in $\phi$
decays 
into 
%
%
%
a
fully anti-symmetric
entangled 
state with $J^{\P\C}=1^{- -}$:
\begin{equation}
  |i \rangle   =  {1\over \sqrt{2}} \left\{ |\kn \rangle |\knb \rangle - 
 |\knb \rangle |\kn \rangle
\right\} 
= {\mathcal{N}\over \sqrt{2}} \left\{ |\ksn \rangle |\kln \rangle - 
 |\kln \rangle |\ksn \rangle
\label{eq:state}
\right\}~,
\end{equation}  
with $\mathcal{N}={\sqrt{(1+|\epsilon_{\rm S}|^2)(1+|\epsilon_{\rm L}|^2)}}/{(1-\epsilon_{\rm S}\epsilon_{\rm L})} \simeq 1$ a normalization factor,
and $\epsilon_{\rm S,L}$ the small \CP impurities in the mixing of the physical states (\ksln) with definite widths 
($\Gamma_{\rm S,L}$)
and masses ($m_{\rm S,L}$).
{It is worth noting that
Bose statistics 
and angular momentum conservation 
forbid the appearance in state (\ref{eq:state}) and its time evolution
of terms with two identical bosons, 
like $\ksn\ksn$ or $\kln\kln$.}
\par
In this paper we focus on the
study of the \CP violating process ${\phi\rightarrow\ksn\kln\rightarrow\pi^+\pi^-\pi^+\pi^-}$
described by the following decay intensity as a function of the kaon decay proper times $t_1$ and $t_2$:
\begin{eqnarray}
\label{eq:intensity}
I(t_1,t_2) &=&
\frac{ | \mathcal{N} |^2 }{2} |\langle \pi^+\pi^- |T| \ksn  \rangle|^4  |\eta_{+-}|^2  \Big\{  e^{-\Gamma_{\rm L} t_1 -\Gamma_{\rm S} t_2}
+ e^{-\Gamma_{\rm S} t_1 -\Gamma_{\rm L} t_2} 
\nonumber \\
& & -2e^{ - \frac{(\Gamma_{\rm S}+\Gamma_{\rm L})}{2} (t_1+t_2)}\cos[ \Delta m (t_1-t_2)]
\Big\}~,
\end{eqnarray}
with $\eta_{+-}=\langle \pi^+\pi^- |T| \kln  \rangle/\langle \pi^+\pi^- |T| \ksn  \rangle$ the ratio of \klsn decay amplitudes into $\pi^+\pi^-$, and $\Delta m = m_{\rm L}-m_{\rm S}$.
The decay intensity (\ref{eq:intensity})
exhibits
a fully destructive quantum interference phenomenon
that prevents both kaons to decay into $\pi^+\pi^-$ at the same time,
even when the two decays are space-like separated events.
This EPR-type correlation
%
constitutes 
a 
very powerful
tool for 
testing
the quantum coherence of state (\ref{eq:state})
at the utmost precision, and to search for possible decoherence effects.
%
%
%
In fact,
the key feature of 
the entangled state (\ref{eq:state}) resides
in its non-separability.
It has been suggested~\cite{bertlmann1}
that, according to Furry's hypothesis~\cite{furry},
soon after the $\phi$-meson {decays}
the 
state  
might spontaneously
factorize to an equally weighted statistical mixture of states
$|\ksn \rangle |\kln \rangle$ or $|\kln \rangle |\ksn \rangle$
(or also --  depending on the decoherence mechanism -- $|\kn \rangle |\knb \rangle$ or $|\knb \rangle |\kn \rangle$),
loosing its coherence.
\par
A way to describe 
such deviations from quantum 
mechanics 
\cite{bertlmann1,eberhard} 
%
%
%
is 
to introduce 
a decoherence parameter $\zeta_{\rm SL}$ 
and a factor
$(1-\zeta_{\rm SL})$ multiplying the interference term in the quantum mechanical expression 
(\ref{eq:intensity}) 
in the $\{\ksn,\kln\}$ basis:
\begin{eqnarray}
\label{eq:intensitydec}
I(t_1,t_2;\zeta_{\rm SL}) &\propto&
e^{-\Gamma_{\rm L} t_1 -\Gamma_{\rm S} t_2}
+ e^{-\Gamma_{\rm S} t_1 -\Gamma_{\rm L} t_2} 
\nonumber \\
& & -2 (1-\zeta_{\rm SL}) e^{ - \frac{(\Gamma_{\rm S}+\Gamma_{\rm L})}{2} (t_1+t_2)}\cos[ \Delta m (t_1-t_2)]
~.
%
\end{eqnarray}  
The case $\zeta_{\rm SL}=0$ corresponds to 
quantum mechanics,
while
for $\zeta_{\rm SL}=1$ 
the case of 
spontaneous factorization of the state
is obtained, i.e. total decoherence.
Different $\zeta_{\rm SL}$ values
correspond to intermediate situations between these two.
{
Analogously,  
for a decoherence mechanism acting in the $\{\kn,\knb\}$ basis~\cite{bertlmann1},
the 
modified 
decay intensity $I(t_1,t_2;\zeta_{0\bar{0}})$ can be defined
by introducing a decoherence parameter $\zeta_{0\bar{0}}$
and a factor $(1-\zeta_{0\bar{0}})$ 
multiplying the interference term in 
this basis (see appendix \ref{appendix}).
}
\par
{
In another phenomenological model \cite{bertlmann2} decoherence is introduced 
at a more fundamental level
via a 
simple 
 dissipative term in the Liouville–von Neumann equation for the density matrix of the state, 
assuming \CP and \CPT invariance.
Decoherence in this model is predicted to become stronger with increasing distance between the kaons,
and is governed by a parameter $\lambda$. This parameter 
is still in simple correspondence with the
decoherence parameter $\zeta_{\rm SL}$ mentioned above 
by $\lambda\simeq \zeta_{\rm SL} \Gamma_{\rm S}$ (see appendix \ref{appendix}).
}
\par
At a microscopic level, in a quantum gravity framework, 
space-time might be
subject
to inherent non-trivial quantum metric and topology 
fluctuations
at the Planck scale (${\sim10^{-33}\hbox{~cm}}$), 
called generically {\it space-time foam}, with associated microscopic event horizons.
This space-time structure
might induce a pure state to evolve into a mixed state, i.e. 
decoherence of apparently isolated matter systems~\cite{hawk2}.
A theorem proves that this kind of decoherence necessarily implies 
\CPT violation, i.e. that the quantum mechanical operator generating \CPT 
transformations is {\it ill-defined}~\cite{wald}.
\par 
The above mentioned decoherence 
mechanism 
due to quantum gravity effects
inspired the formulation of a phenomenological model consistent with this hypothesis~\cite{ellis1,ellis2},
in which 
a single kaon
is 
described by a 
density matrix $\rho$ that
obeys  a specifically modified Liouville-von Neumann equation:
\begin{equation}
  \label{eq:evolmod2}
  \frac{\partial \rho}{\partial t} = -i{\bf H}\rho  +i\rho {\bf H}^{\dagger} + 
{\bf L}(\rho; \alpha,\beta,\gamma)~,
\end{equation}
where 
${\bf H}$ is the effective Hamiltonian 
describing the kaon system {(not assuming 
\T, 
\CP or \CPT invariance)}, 
and
the extra term ${\bf L}(\rho; \alpha,\beta,\gamma)$
induces decoherence in the system, and depends on
%
three real parameters, $\alpha, \beta$ and $\gamma$,
which violate \CPT symmetry and quantum mechanics (and
satisfy the 
relations
$\alpha$, $\gamma >0$ and $\alpha \gamma > \beta^2$).
%
They have mass dimension and 
are supposed to be
at most of $\mathcal {O} (m^2_{\rm K}/M_{\rm{Planck}}) \sim 2 \times 10^{-20} \,\mbox{GeV}$~\cite{ellisstring92,ellis2000},
where $M_{\rm{Planck}}=1/\sqrt{G_N}= 1.22\times 10^{19} \mbox{~GeV}$ is the Planck mass.
In the entangled kaon system at a $\phi\hbox{-factory}$~\cite{peskin}
this decoherence model can be tested in the channel
$\phi\rightarrow \ksn\kln \rightarrow \pi^+\pi^-\pi^+ \pi^-$
and in the simplifying hypothesis of complete positivity~\cite{benattiall},
 i.e. $\alpha=\gamma$ and
$\beta=0$, with $\gamma$ as the 
parameter describing the phenomenon
through the corresponding decay intensity $I(t_1,t_2;\gamma)$ {-- see appendix \ref{appendix}.}
\par
As 
mentioned
above, 
in a quantum gravity 
framework inducing decoherence, the \CPT operator is {\it ill-defined}. 
This might have
a consequence in correlated neutral kaon states, where
the resulting loss of particle--antiparticle identity could induce a 
breakdown of the correlation of state (\ref{eq:state}) 
imposed by Bose statistics 
\cite{mavro1,mavro2,mavro3}.
As a result the initial state (\ref{eq:state}) can acquire a small 
symmetric $\kn\knb$
component and be parametrized 
as:
\begin{eqnarray}
|i \rangle  &=&  {1\over \sqrt{2}} \left[ |\kn\rangle |\knb\rangle -|\knb\rangle |\kn \rangle  
+ \omega \left( |\kn\rangle |\knb\rangle + 
|\knb\rangle |\kn\rangle \right) \right] \nonumber\\
&\propto &
\left[ |\ksn\rangle |\kln\rangle -|\kln\rangle |\ksn \rangle  
+ \omega \left( |\ksn\rangle |\ksn\rangle -
|\kln\rangle |\kln\rangle \right) \right] 
\label{eq:state5}
~,
\end{eqnarray}  
where $\omega
=|\omega| e^{i\phi_{\omega}}$ is a complex parameter describing this particular \CPT
violation phenomenon, and $I(t_1,t_2;\omega$) the corresponding modified decay intensity {(see appendix \ref{appendix})}.
The order of magnitude of $\omega$
is expected to
 be at most $|\omega| \sim \left [(m^2_{\rm K}/M_{\rm{Planck}})/\Delta \Gamma \right ]^{1/2} \sim 10^{-3}$ with $\Delta \Gamma = \Gamma_{\rm S} - 
\Gamma_{\rm L}$. 
\\
As an ancillary consideration, 
from the 
measurement of $\omega$ 
one 
can 
extract 
the ratio of the 
branching ratios of the $\phi$--meson decay into a symmetric/anti-symmetric 
$\kn\knb$ state:
\begin{eqnarray}
\label{eq:brksks}
|\omega|^2=
\frac{\rm{BR}( \phi\rightarrow\ksn\ksn
,\kln\kln
)}{\rm{BR}( \phi\rightarrow \ksn\kln)}
~,
\end{eqnarray}
where
$\rm{BR}( \phi\rightarrow\ksn\ksn,\kln\kln)$ is intended as the branching fraction of the $\phi$ decay
into a $\ksn\ksn$ or $\kln\kln$ pair\footnote{{
The \C-even $\kn\knb$ background produced in two photon processes
or in $f_0$,$a_0$ decays
is very small in general and can be considered negligible
also in this context \cite{dunietz,kloekkg}.
It would be anyhow experimentally distinguishable 
from the $\omega$-effect due to different kinematics~\cite{kloekkg} or $\sqrt{s}$
dependence across the $\phi$-resonance~\cite{didohand,mavro1}. 
}}, {and $\rm{BR}( \phi\rightarrow\ksn\kln)$ is the known branching fraction
into a $\ksn\kln$ pair in state (\ref{eq:state})~\cite{pdg}.}
%
\par
The theoretical decay times distributions 
of 
the phenomenological models mentioned above, after integration on the sum $(t_1+t_2)$,
are used to fit the experimental 
distribution
obtained with the
KLOE 
experiment 
at DA$\Phi$NE 
as a function of the absolute difference of times $\Delta t=|t_1-t_2|$.
In the following,
 the experimental set-up is briefly described in section~\ref{sec:kloe}. 
The event selection (section~\ref{sec:selection}), the residual background  evaluation due to the 
non-resonant $\pi^+\pi^-\pi^+\pi^-$
 production and kaon regeneration processes (section~\ref{sec:BkgEval}), and the evaluation of efficiency as a function of $\Delta t$  (section~\ref{sec:effi}) are then presented.
The adopted fit procedure 
is detailed in section~\ref{sec:fit}. 
After the study of the different sources of systematic uncertainties in section~\ref{sec:syst},
the final results on the various decoherence and  \CPT-violating parameters are presented 
in section~\ref{sec:results}.
%
\par
These results 
are obtained from the analysis of 
a data sample
larger by about a factor four, statistically independent, and with an improved 
background evaluation and signal selection criteria
with respect to previous KLOE analyses \cite{kloeqm2006}.
They 
are complementary 
 to the 
 results 
 on the test of \CPT and Lorentz symmetry obtained studying the same process~\cite{kloecpt2013}.

\section{The KLOE detector at DA$\Phi$NE}\label{sec:kloe}
The data 
were
collected with the KLOE detector at the  DA$\Phi$NE $e^+e^-$ collider 
{\cite{dafne1,dafne2,dafne3}},
that 
operates
at a center-of-mass energy 
{corresponding to
the mass of the $\phi$ meson, i.e. $1019.4$~MeV}. Positron and electron beams of equal energy collide at an angle of 
${(\pi-0.025)}$ rad, producing $\phi$ mesons with a small 
momentum
in the horizontal plane, 
$p_{\phi}\simeq 13 \hbox{~MeV}$, 
and decaying
about
 $34\%$
 of the time
into nearly collinear 
\ksnkln
pairs.
The data sample analyzed in the present work 
corresponds to an integrated luminosity of 
{about
$1.7$~fb$^{-1}$, 
i.e. 
to 
about
$ 1.7 \times 10^9$ 
$\phi\rightarrow\ksnkln$
decays 
produced.}
\par
The beam pipe at the interaction region of 
DA$\Phi$NE has a spherical shape, with a radius of 10 cm, and is made of a 
62\% beryllium--38\% aluminum alloy, 500 $\mu$m 
thick. A thin beryllium cylinder, 50 $\mu$m thick, with 4.4 cm radius, and coaxial with the beam, ensures electrical continuity.
\par
The 
detector consists of a large cylindrical drift chamber (DC), surrounded by a lead/scintillating-fiber sampling calorimeter (EMC). 
A superconducting coil surrounding the calorimeter provides a 0.52 T magnetic field. 
%
%
The 
DC \cite{dc} is 4~m in diameter and 3.3~m long. The chamber 
shell is made of carbon-fiber/epoxy composite, and the gas used is a 90\% helium--10\% isobutane mixture. 
These features maximize transparency to photons and reduce $\kln\rightarrow \ksn$ regeneration and multiple scattering. 
 The spatial resolution is 
 {
 $\sigma_{xy}\simeq 150~\mu\mathrm{m}$ and $\sigma_z\simeq 2$~mm}
 in the transverse and longitudinal projections, respectively.
Vertices are reconstructed with a spatial resolution 
of 
$3$~mm.
The momentum resolution is $\sigma(p_{\perp})/p_{\perp}=0.4\%$,
and the $\ksn\rightarrow \pi^+\pi^-$ invariant mass is reconstructed with a resolution of 
$1 \hbox{~MeV}$.
The calorimeter \cite{emc} is divided into a barrel and two endcaps, covering 
$98\%$
of the solid angle. 
The modules are read out at both ends by photomultiplier tubes 
with a read-out granularity of 
about $4.4 \times 4.4 \hbox{~cm}^2$. 
The arrival times of particles and the three-dimensional positions
 of the energy deposits are determined from the signals at the two ends;
 fired {cells} close in space and time are grouped into an 
 energy
 cluster. For each cluster, the energy 
 $E$
 is the sum of the cell energies,
 the time 
 $t$
 and the position 
 $r$
 are 
 calculated as
  energy-weighted averages over the fired
 cells. The energy and time resolutions are 
 $\sigma_E/E =5.7\% /\sqrt{E(\hbox{GeV})}$
 and
 $\sigma_t =54\hbox{~ps} /\sqrt{E(\hbox{GeV})}  \oplus 100 \hbox{~ps}$, respectively.
\par
The trigger \cite{trigger} 
uses a two level scheme. The first level trigger is a fast trigger with a minimal delay which starts the acquisition of the EMC front-end-electronics. 
The second level trigger is based on the energy deposits in the EMC (at least 50 MeV in the barrel and 150 MeV in the end-caps) 
or on the hit multiplicity information from the DC. The trigger conditions are chosen to minimise the machine background, 
and recognise Bhabha scattering or cosmic-ray events. Both the calorimeter and drift chamber triggers are used for recording 
interesting events.
\par
The response of the detector to the decays of interest and the various backgrounds are studied by using the KLOE Monte Carlo (MC) simulation program \cite{offline}. 
Changes in the machine operation and background conditions are taken into account.
The MC samples used in the present analysis amount to an equivalent integrated luminosity of 17~fb$^{-1}$ for the 
signal, and to 3.4~fb$^{-1}$ for all
{main}
 $\phi$ decay channels.


\section {Event selection}\label{sec:selection}
The following scheme is adopted for the event sample selection.
Candidate signal events are topologically identified
requiring the reconstruction of two vertices with two 
opposite curvature tracks each.  At least one vertex is required within a cylindrical fiducial volume ($\rho=\sqrt{x^2+y^2}<10~\hbox{cm}$ and $|z|< 20~\hbox{cm}$) 
centered at the collision point. The latter
is evaluated 
run-by-run from Bhabha scattering events.
For vertex $i$ $(i=1,2)$ 
the kaon momentum $\pvec_i$ and energy $E_i$ are evaluated from the 
pion
momenta 
{
$\pvec_{i+}$ and $\pvec_{i-}$
as
${\pvec_{i}=\pvec_{i+}+\pvec_{i-}}$ and ${E_{i}=\sqrt{|\pvec_{i}|^2+m_{\knped}^2}}$.
Afterwards}
the following selection criteria are applied (preselection):
%
%
%
%
%
\begin{itemize}
\item 
 $|m_i(\pi^+\pi^-)-m_{\knped}|<5~\hbox{MeV}$, with $m_i(\pi^+\pi^-)$ the invariant mass calculated from 
 {
 $\pvec_{i\pm}$
 and assuming}
 the charged pion mass hypothesis;
\item
$ \Big| |\pvec_{i+}^{~*}+\pvec_{i-}^{~*}|
-p^{~*}_{\knped}\Big|< 10~\hbox{ MeV}~,$
 with $\pvec_{i \pm}^{~*}$ 
 the pion momenta of vertex $i$ calculated in the $\Pphi$ rest frame,
 $p^*_{\knped}=\sqrt{\frac{s}{4}-{m_{\knped}}^2}$ 
 the kaon momentum calculated from
the kinematics of the $\f\to\ksn\kln$ decay, where \sqrts, the $e^+e^-$ C.M. energy,
is obtained run-by-run from Bhabha scattering events;
\item  $-50$ 
$< E^2_{\rm miss}  -|\pvec_{\rm miss}|^2<$ 10 MeV$^2$~,
with
$\pvec_{\rm miss} = \pvec_\Pphi -\pvec_{1}-\pvec_{2}$
and $ E_{\rm miss} =  E_\Pphi - E_{1} - E_{2}$~;
\item  $\sqrt{E^2_{\rm miss}+|\pvec_{\rm miss}|^2} <10$ MeV~.
\end{itemize}
\par
A kinematic fit
is 
then
performed in order to improve the resolution 
on the kaon decay vertices.
 The two decay vertices are 
 parametrized as:
%
 \begin{eqnarray}
 \vec{V}_{i} = \vec{V}_{\phi} + \lambda_{i} \, \hat{n}_{i}~,
 \end{eqnarray}
 where $\vec{V}_{i}$ are the positions of the vertices,
 $\vec{V}_{\phi}$ is the $\phi$ decay position, $\lambda_{i}$ are the decay
 lengths, and $\hat{n}_{i}=\pvec_{i}/|\pvec_{i}|$ are unit vectors identifying the kaon
 directions as reconstructed from tracks.
A global kinematic fit solves for $\lambda_{i}$
and $\vec{V}_{\phi}$ maximizing the log-likelihood function:
\begin{equation}
    \ln L =  \sum_{i=1,2}\ln P_i(\vec{V}_{i}^{(\mathrm{rec})} - \vec{V}_i^{(\mathrm{fit})} ) + \ln P_{\phi}(\vec{V}_{\phi}^{(\mathrm{rec})} - 
    \vec{V}_{\phi}^{(\mathrm{fit})} ) 
    ~,
\end{equation}
where $P_i$ and $P_{\phi}$ are the probability density functions representing the resolutions for $\vec{V}_{i}$
and $\vec{V}_{\phi}$,
as obtained from MC.
 As the main uncertainty in the  vertex position is in the direction orthogonal
 to the kaon line of flight (due to the large opening angle of the two pions), 
 the kinematic fit 
 takes into account separately the
 vertex resolution projected along this direction and in the transverse plane.
All events with $-2 \ln L < 30$ and
{
$\frac{| \lambda_{i}^{(\mathrm{fit})} - \lambda_{i}^{(\mathrm{rec})} |}{ \sigma(\lambda_{i}^{(\mathrm{fit}) } )}< 10$
are retained.}
%
%
%
From 
the decay lengths $\lambda_{i}$ the proper times are evaluated: 
{$t_{i} =\frac{\lambda_{i}}{ | \pvec_{i} | }m_{\knped} $~.}
\par
The resolution 
on the difference 
$\Delta t=|t_2-t_1|$ 
is strongly correlated with
the opening angle $\theta_{\pi\pi}$ of the pion tracks,
and is worsening for large values of $\theta_{\pi\pi}$.
%
A final selection cut is therefore applied to
candidate events, requiring both vertices with $\cos(\theta_{\pi\pi}) > -0.975$, obtaining a further 
improvement on the $\Delta t$ resolution with a moderate loss in efficiency.
The core width of the $\Delta t$ resolution at the end of the selection is 
approximately
$0.7~\tau_{\rm S}$.
\section{Background evaluation}\label{sec:BkgEval}
There are two main background sources after the selection for the signal described above:
the non-resonant
 production of four pions, $e^+e^- \to \pi^+\pi^-\pi^+\pi^-$, and kaon regeneration on the beam pipe.
The remaining background due to semileptonic \kln decays can be considered negligible, 
being 
uniformly distributed in $\Delta t$, and 
amounting
in total from MC 
to less than 
{
$0.2\%$}
in the range ${0< \Delta t < 12~\tau_{\rm S}}$.
\par
\begin{figure}[hbt!]
     \centering
     \includegraphics[scale = 0.37]{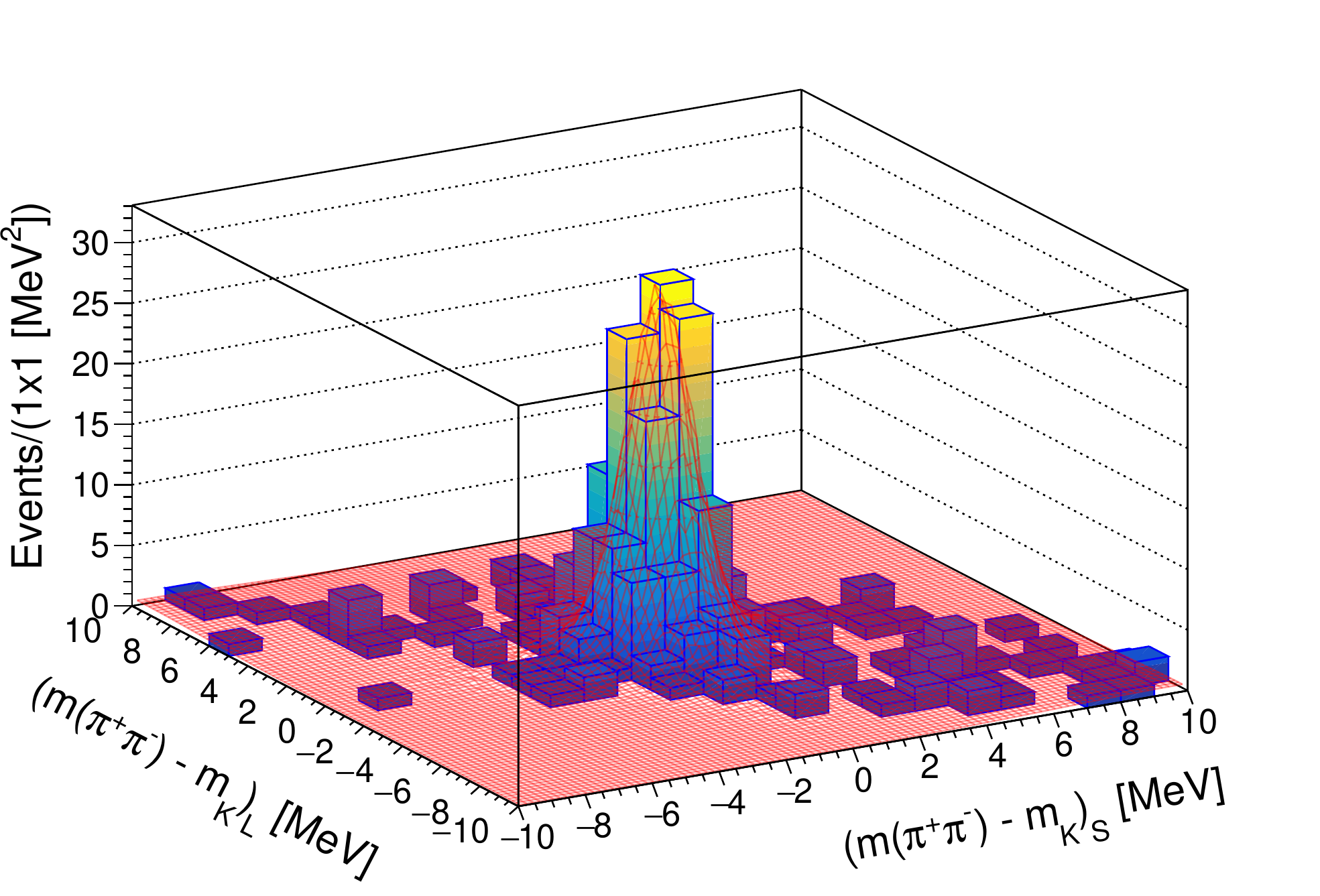}
     \includegraphics[scale = 0.37]{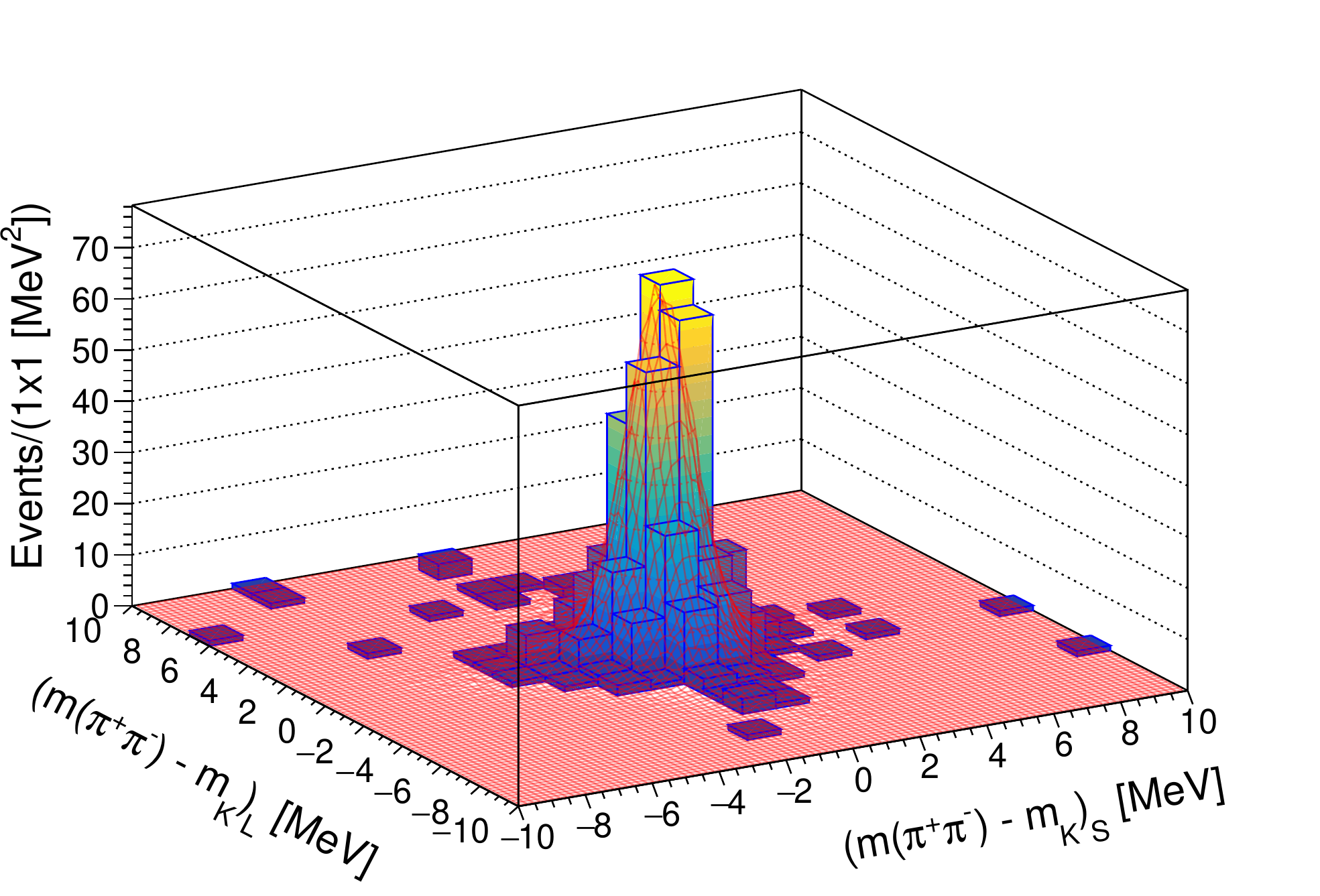}
     \includegraphics[scale = 0.37]{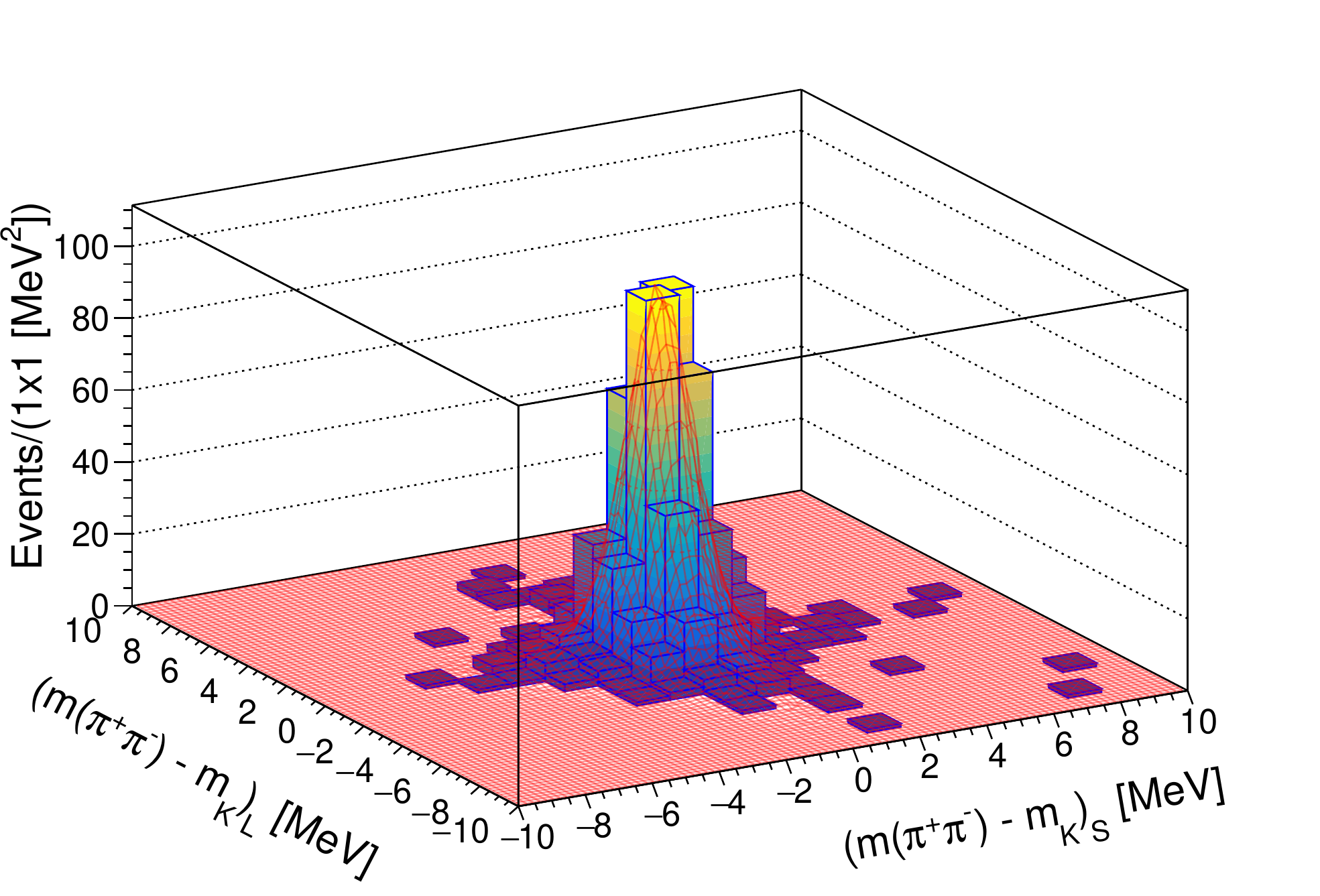}
     \includegraphics[scale = 0.37]{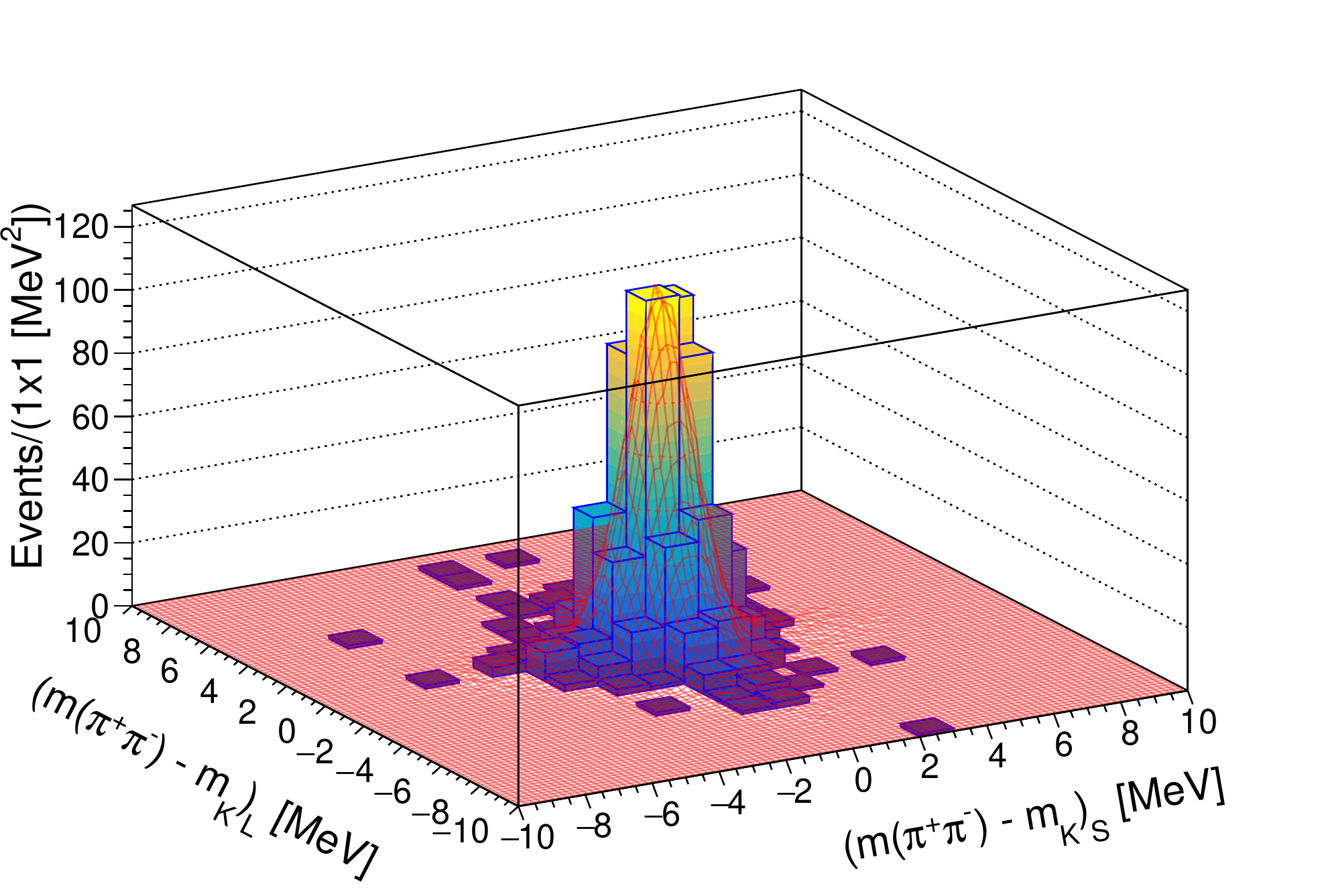}
     \caption{{
     Invariant mass distribution of \kln vs \ksn vertices for four $\Delta t$ bins: $0< \Delta t <1~\tau_{\rm S}$ (top left),
     $1< \Delta t <2~\tau_{\rm S}$ (top right), $2< \Delta t <3~\tau_{\rm S}$ (bottom left), and $3< \Delta t <4~\tau_{\rm S}$ (bottom right).
      The histograms of data have superimposed the result of the unbinned fit (red).  
}
     }
     \label{fig:2dminv1stbin}
 \end{figure}

The process $e^+e^- \to \pi^+\pi^-\pi^+\pi^-$, 
although not dominant ({about $0.5\%$} in the range ${0< \Delta t < 12~\tau_{\rm S}}$), 
is  concentrated at 
$\Delta t\approx 0$, which is the most sensitive region to decoherence effects described in section~\ref{sec:intro}.
This background
is evaluated
by studying the 
two-dimensional invariant mass distribution of the reconstructed kaon decay vertices\footnote{
In the following we conventionally name \ksn (\kln) the kaon with its reconstructed decay
closest to
(farthest from) 
the $\phi$ production point,
even though actual \ksn and \kln decays close in time would be quantum mechanically 
indistinguishable due to the overlap $\langle\ksn|\kln\rangle \neq 0$ originated by \CP violation.
}
 in bins of $\Delta t$.
In the distribution corresponding to the first bin ($0<\Delta t < 1~\tau_{\rm S}$) shown in figure \ref{fig:2dminv1stbin} 
the signal peak at the center and the background contribution
distributed along the second diagonal (due to a correlation introduced by the selection and by kinematical constraints) can be clearly identified.
An unbinned maximum likelihood fit is performed
in order to evaluate the number of background events. 
The model used to fit the observed distribution is:
\begin{eqnarray}
N_{\rm obs}(x,y) = N_S \cdot S(x,y) + N_B \cdot G(z; \mu_z=0 , \sigma_B)
~,
\end{eqnarray}
with $x,y = \left[ m(\pi^+\pi^-)-m_{\knped} \right]_{\rm S,L}$ the invariant mass shift for reconstructed \ksln vertices, 
$S(x,y)$ the template for the signal shape obtained from MC after the selection, 
not imposing
the invariant mass constraint, 
$G(z; \mu_z=0 , \sigma_B)$ a Gaussian distribution as a function of the variable $z=x+y$ with 
zero mean ($x= -y$) and standard deviation $\sigma_B$ (free parameter in the fit) to model the background along the second diagonal,
$N_S$ and $N_B$ two normalization factors for signal and background, respectively.
The background events are then extrapolated from the $|x,y|<10~\mathrm{MeV}$ region to the nominal 
signal
region $|x,y|<5~\mathrm{MeV}$.
The result is
$n_{B,1}=39 \pm 5$ events for the first bin ($0 \leq \Delta t <1~\tau_{\rm S}$). 
The consistency of this result is checked against different extrapolating regions, from $|x,y|<6~\mathrm{MeV}$ to $|x,y|<10~\mathrm{MeV}$.
The fit is independently repeated for the ($1 < \Delta t <2~\tau_{\rm S}$), ($2 < \Delta t <3~\tau_{\rm S}$), and
($3 < \Delta t <4~\tau_{\rm S}$) bins yielding as result
$n_{B,2}=6 \pm 2$, $n_{B,3}=6\pm 2$, and $n_{B,4}=3\pm 1$ events, respectively.
From these results we conclude that the $e^+e^- \to \pi^+\pi^-\pi^+\pi^-$ background can be considered negligible for $\Delta t > 3~\tau_{\rm S}$.
\par
The regeneration background peaks in the region at  
{$\Delta t \approx 17~\tau_{\rm S}$}
due to the spherical beam pipe \cite{kloeqm2006}.
In the present analysis the fit to the $\Delta t$ distribution 
is restricted 
in the range $0< \Delta t <12~\tau_{\rm S}$ to 
avoid
this background, while keeping the maximum sensitivity to the various decoherence and \CPT violation parameters, which is mainly 
in the region $\Delta t < 6~\tau_{\rm S}$. The remaining regeneration background
is due to the small 
contribution from the thin beryllium cylinder,
largely dominated by the incoherent over the coherent 
regeneration
\cite{qmadd,baldini}. 
The 
latter 
can be therefore neglected.
The $\Delta t$ distribution 
of the 
residual 
incoherent regeneration background 
is spread 
out 
over the region $5< \Delta t < 12~\tau_{\rm S}$,
as an effect
of the cylindrical geometry, and is evaluated with Monte Carlo. 
The $\ksn$ regeneration probability,
$P_{\rm reg}^{\rm cyl}$,
is extrapolated from the measured regeneration 
probability
%
on the spherical beam pipe~\cite{kloeqm2006}, $P_{\rm reg}^{\rm sph}$, knowing the incoherent regeneration 
{cross-section} ratio for beryllium and aluminum,
$r = \frac{\sigma_{\rm inc}(\rm Be)}{\sigma_{\rm inc}(\rm Al)}$, and using the relation:
\begin{equation}\label{eq:rege}
P_{\rm reg}^{\rm cyl} = D_{\rm geom}\cdot P_{\rm reg}^{\rm sph}\cdot f(r)
~,
\end{equation}
where $D_{\rm geom}$ is a factor depending on the beam pipe geometry, 
and 
$f(r)$ a function 
of
the ratio $r$:
\begin{eqnarray}
f(r) = \frac{(\rho_{\rm Be}/A_{\rm Be})\cdot r}{\rho_{\rm AlBe}\left(\frac{w_{\rm Be}}{A_{\rm Be}}\cdot r + \frac{w_{\rm Al}}{A_{\rm Al}}\right)}~,
\end{eqnarray} 
with $\rho_{\rm AlBe}$
 the density of the spherical beam pipe, 
$w_{\rm Be}$
and $w_{\rm Al}$
the $\rm Be/Al$ proportion by weight.
The 
large 
uncertainty on $r$
(from ref.~\cite{baldini} $r\approx 2$, while a preliminary KLOE measurement  
yields $r\approx 0.3$ \cite{iza}) reflects on $f(r)$, which is anyhow smoothly varying with $r$ reaching a plateau for $r\gtrsim 2$,
 and dominates the uncertainty on $P_{\rm reg}^{\rm cyl}$.
%
This quantity is used as input to the MC evaluation of the true $\Delta t$ distribution with and without the effect of regeneration. 
The ratio between the two corresponding 
histograms provides the correction factors $R_j$ to account for regeneration background as a function of $\Delta t$ in the 
fit procedure described in section~\ref{sec:fit}.

\section{Efficiency evaluation}\label{sec:effi}

\begin{figure}[hbt!]
    \centering
    \includegraphics[scale = 0.6]{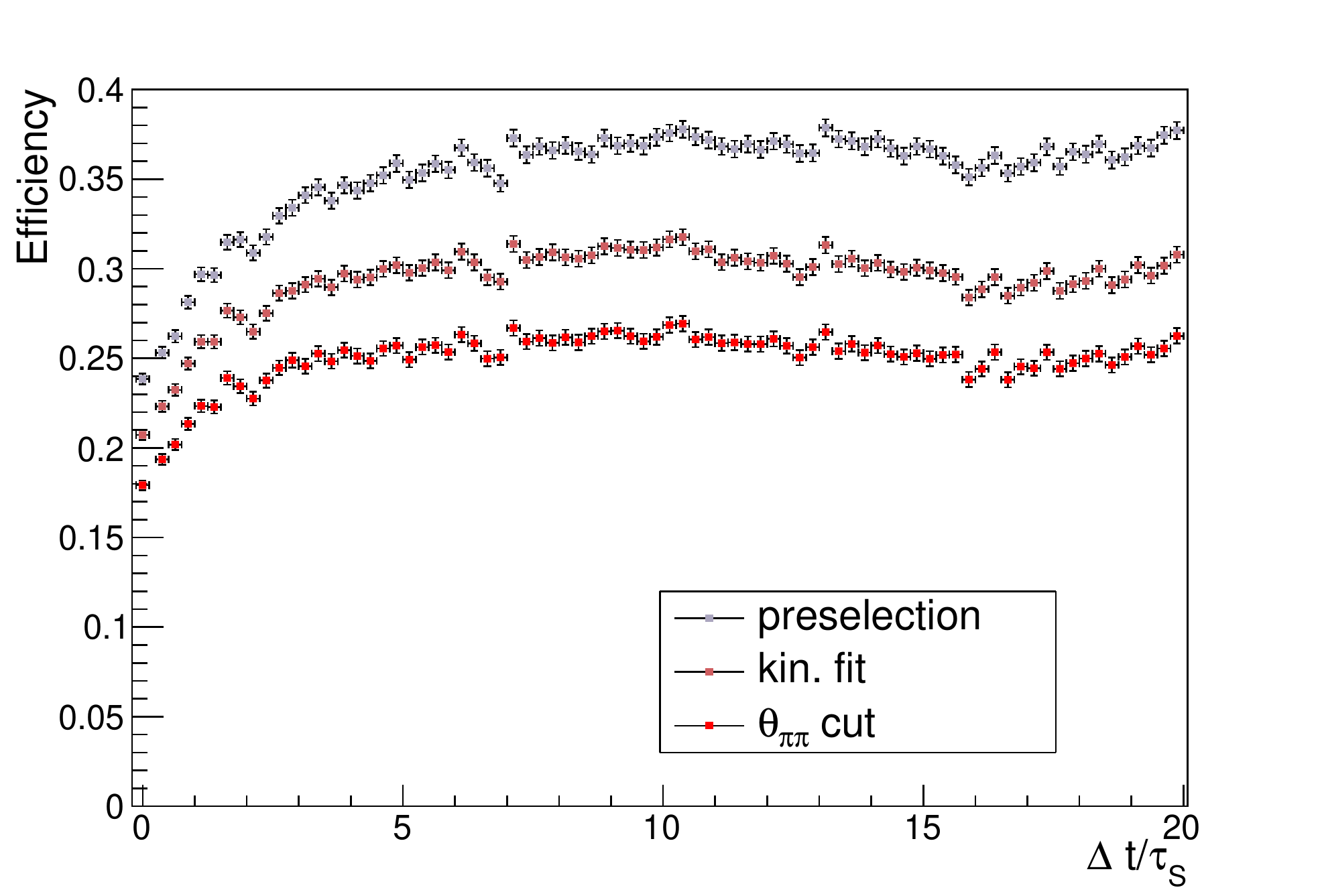}
    \caption{Efficiency evaluated from the Monte Carlo, after the preselection criteria 
    applied (grey), the kinematic fit (brown), and the cut on the $\theta_{\pi\pi}$ angle (red). {The vertical error bars indicate the statistical uncertainty
    of the Monte Carlo sample}.}
    \label{fig:effi}
\end{figure}

The total selection efficiency can be parametrized as the product:
\begin{equation}
   \epsilon_{\rm tot} = \epsilon_{\rm trig}~\epsilon_{\rm reco}~\epsilon_{\rm cuts}
   ~,
\end{equation}
with $\epsilon_{\rm trig}$ the efficiency due to the trigger, $\epsilon_{\rm reco}$ the efficiency of the reconstruction procedure and $\epsilon_{\rm cuts}$ the efficiency due to the selection cuts. The latter is derived from the MC, while 
the MC efficiencies of
the first two are corrected
using data from an
independent control sample, as described in the following. 
It has to be underlined 
that for the present analysis only the dependence of $\epsilon_{t\rm ot}$ on $\Delta t$ is crucial, while its absolute global value is 
not relevant.
 The efficiency 
$\epsilon_{\rm tot}$ as obtained from MC
 is shown in figure~\ref{fig:effi} as a function of $\Delta t$, before and after the application of the kinematic fit and the cut on the opening angle. 
The efficiency is in average  $\sim 25\%$ 
with a small 
reduction
at $\Delta t \approx 0$ that is due to two concurrent effects:
    (i) longer extrapolation length for both tracks 
    originating in
    the interaction point that enhances the probability to fail the reconstruction;
    (ii) the possible swap of tracks associated to two different kaon decay vertices,
when the two vertices are close in time $\Delta t \approx 0$.
\par
\begin{figure}[hbt!]
    \centering
    \includegraphics[scale = 0.6]{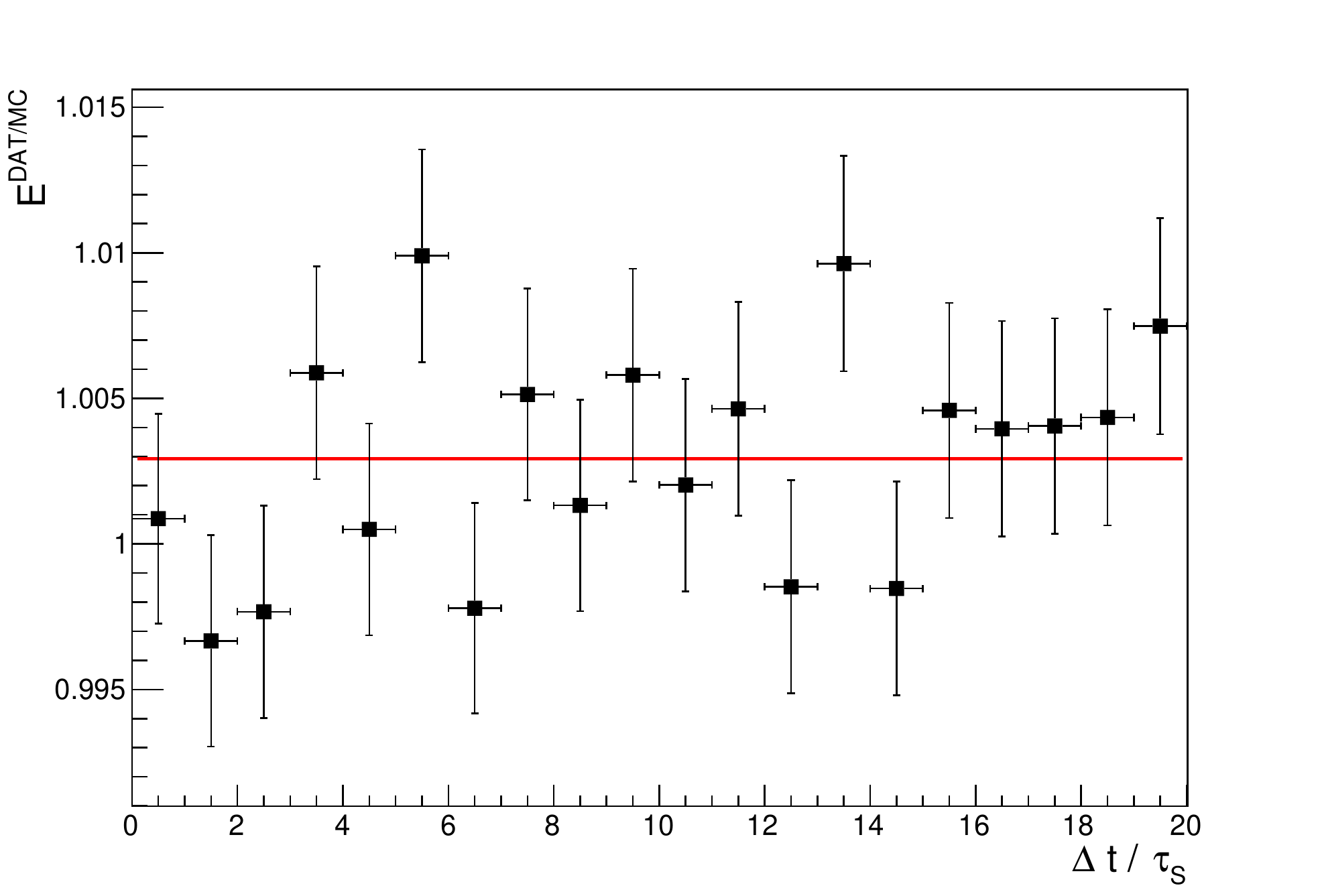}
    \caption{Data/MC efficiency correction as a function of $\Delta t$. 
    The 
    superimposed line is the result of a fit with a constant. }
    \label{fig:datMCCorrection}
\end{figure}

The trigger and reconstruction efficiencies provided by the MC are checked with data, using an independent control sample of 
$\ksn\kln \rightarrow \pi^+\pi^-\pi\mu\nu$ events, selected to have 
high
purity and to have an overlap with the momentum distribution of the signal. 
The applied selection criteria 
ensure 
the statistical independence of the control sample from the signal and
a purity 
of $95\%$, with the residual background dominated by the $\ksn\kln \rightarrow \pi^+\pi^- \pi e \nu$ decay.
The efficiency correction has been evaluated as the ratio between data and MC $\Delta t$ distributions
of $\ksn\kln \rightarrow \pi^+\pi^- \pi \mu \nu$ events.
A fit with a constant indicates a quite small average correction, as shown in figure~\ref{fig:datMCCorrection}.

\section{Fit}\label{sec:fit}
\begin{figure}[hbt!]
\centering
    \includegraphics[scale = 0.6]{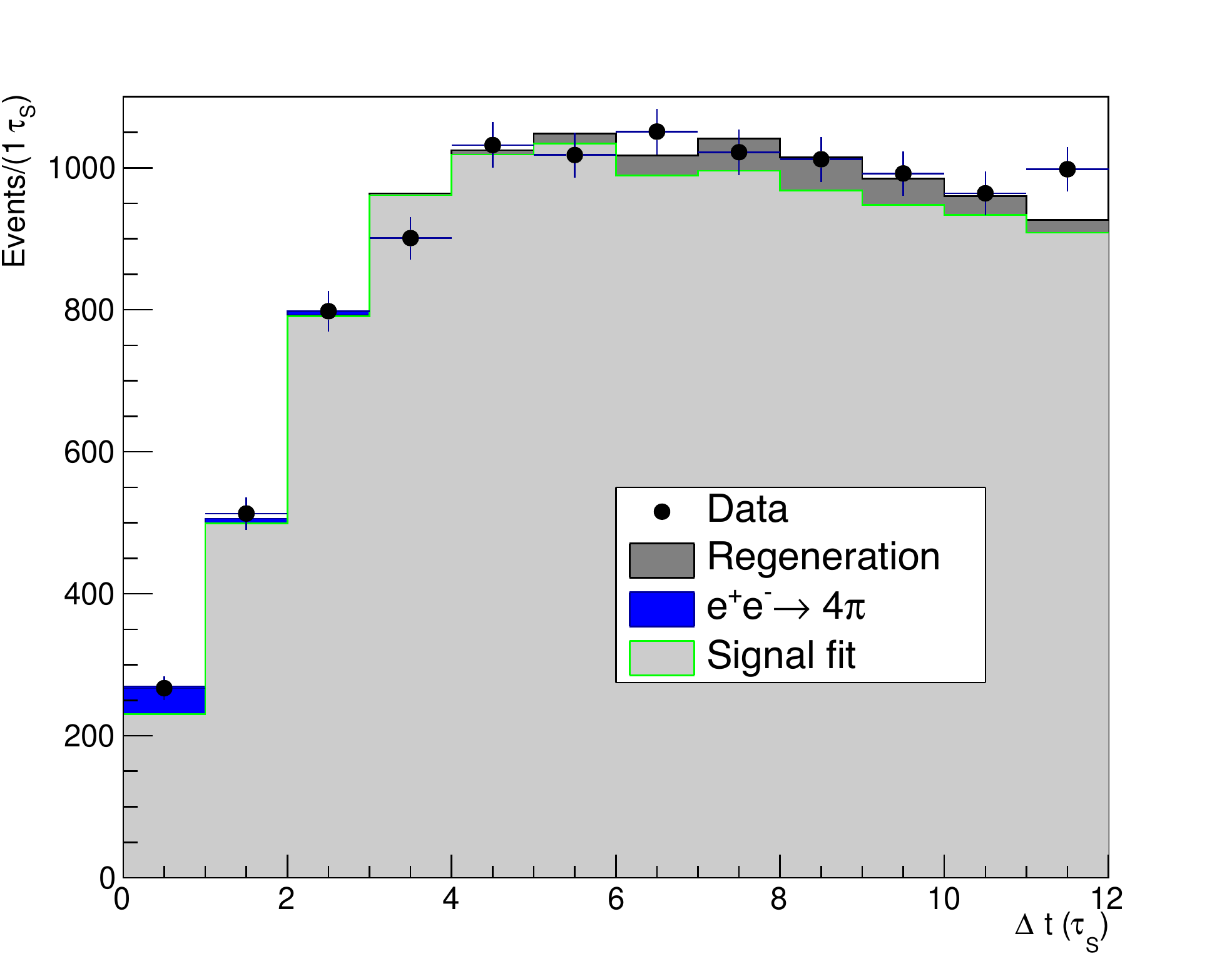}
    \caption{{Data and 
    fit distribution}
    for the $\zeta_{\rm SL}$ decoherence model with background contributions displayed.}
    \label{fig:Fit_SL}
\end{figure}

As already pointed out in section~\ref{sec:BkgEval}, the fit is performed on the measured $\Delta t$ distribution
in the range 
$ 0 < \Delta t< 12~\tau_{\rm S}$, in order to minimize the regeneration background.
The bin width of the data histogram is $\overline{\Delta t}=1~\tau_{\rm S}$, of the same size of the $\Delta t$ resolution.
The fit function in the $j$-th true $\Delta t$ bin
is evaluated 
integrating the decay intensities 
described in section~\ref{sec:intro}
over the sum $(t_1+t_2)$,
 at fixed $\Delta t$,
and over the bin width:
%
\begin{equation}
I({\bf q})_j^{\rm th} = \int_{(j-1)\overline{\Delta t}}^{j \overline{\Delta t}}d(\Delta t)\int_{\Delta t}^{\infty} 
I(t_1, t_2|{\bf q}) ~d(t_1+t_2) ~,
\end{equation}
with ${\bf q}$ the vector of the decoherence and \CPT violation parameters.
Then the following expression
for the expected number of events in the $i$-th reconstructed $\Delta t$ bin
is used to fit the observed $\Delta t$ distribution:
\begin{equation}
N({\bf q})_i^{\rm th} =\mathcal{N} E^{\frac{\rm DAT}{\rm MC}}_i\sum_{j}S_{i,j}\epsilon_j R_j I({\bf q})_j^{\rm th}
\end{equation}
with 
$E_i^{\frac{\rm DAT}{\rm MC}}$ the data/MC efficiency correction discussed in section \ref{sec:effi}, $S_{i,j}$ the smearing matrix describing the $\Delta t$ resolution obtained from MC, $\epsilon_j$ the MC efficiency, $R_j$ a factor
describing the regeneration contribution
 according to (\ref{eq:rege}) and taking into account its $\Delta t$ shape, 
and $\mathcal{N}$ a global normalization factor.
Finally the $4\pi$ background evaluated as described in section~\ref{sec:BkgEval} is subtracted from the data histogram,
{with $N_i^{\rm data}$ the resulting number of observed events in the $i$-th bin.}
The following $\chi^2$ function is minimized by the fit:
\begin{equation}
    \chi^2 = \sum_{i}\left(\frac{N_i^{\rm data} - N({\bf q})_i^{\rm th} }{\sigma_i}\right)^2 ~,
\end{equation}
with 
$\sigma_i$ evaluated by summing in quadrature the statistical uncertainty of data and the subtracted $4\pi$ background, and 
including the uncertainties for MC efficiency and data/MC corrections.
The measured $\Delta t$ distribution and, as an example, the result of the fit for the $\zeta_{\rm SL}$ decoherence model, are shown in figure \ref{fig:Fit_SL};
{the total number of signal events from this fit 
is $10278\pm105$.}
\section{Systematic uncertainties}\label{sec:syst}

In order to evaluate the systematic uncertainty on the fit results,
the whole fitting procedure 
is repeated varying the selection cuts, the number of background events,
the $\Delta t$ resolution,
and the physical constants given in input to the theoretical models, as described in the following. 

\begin{itemize}
\item
{\bf Selection {Cut} Stability} --
The selection cuts 
are
all varied in order to test the MC 
in evaluating
the corresponding efficiency variation. 
The cuts 
are varied according to their resolution $\sigma$ in steps of $\pm 1\sigma$ and $\pm2\sigma$ to check the stability of the results. 
For the evaluation of the systematic uncertainty only the $\pm 1 \sigma$ variation 
is considered. The uncertainties from the various cuts 
are added in quadrature, with the exception of 
the uncertainty due to the variation of the cut on the invariant mass of the two vertices, which 
is not considered. In fact the latter is strongly correlated 
with 
the variation of the $4\pi$ background contamination, which 
is the dominant effect in this case, and
has to be correspondingly rescaled for each invariant mass cut (see section~\ref{sec:BkgEval}).
Therefore 
the effect of the invariant mass cut variation 
is 
automatically included in the evaluation of the systematic uncertainty 
of 
the $4\pi$ background described below.
\item
{\bf $\mathbf{ 4\pi}$ Background} -- 
\label{sec:sys4pi}
To vary the number of background events originated from the non-resonant pion production $e^+e^-\rightarrow \pi^+\pi^-\pi^+\pi^-$ process, the fit procedure 
is 
repeated 
rescaling the $n_{B,i}$  {background yield} according to the uncertainty of the total $4\pi$ background contribution $n_B=n_{B,1}+n_{B,2}+n_{B,3}=51.0\pm5.7$.
%
%
%
\item
{\bf Regeneration} -- 
The factors $R_j$ describing the regeneration background as a function of $\Delta t$ 
 are
varied according to the 
{
   $ R^{\prime}_j = 
   R_j+a\,(R_j-1)$
relation
(no regeneration corresponds to $R_j=1$),}
where 
$a$ parametrizes the fractional variation due to the uncertainty on the regeneration process. 
The latter is dominated by the uncertainty on the knowledge of
the cross-section ratio $r$ through the function $f(r)$, as discussed in section~\ref{sec:BkgEval}.  
A variation of $r$ in the range $\sim 0.3 < r < \infty $ 
corresponds to a
fractional variation 
in the range $- 35\% < a < +10\%$,  which is therefore
considered 
for the evaluation of the systematic uncertainty due to regeneration.
\item
{\bf $\mathbf{ \Delta t}$ Resolution} --
In order to evaluate the MC reliability on the $\Delta t$ resolution description, the proper decay time ($t_{\rm S}$) distribution of $\ksn\rightarrow \pi^+\pi^-$ 
events 
is obtained selecting a sample
relaxing
 the boundary condition $\lambda_{\rm S} > 0$ in the kinematic fit procedure, and requiring 
 for
 the \kln 
 a decay path of at least $12 \mbox{~cm}$.
 This allows to compare the negative tail of the 
 $t_{\rm S}$
 distribution for data and MC, which corresponds to those events in which  the \ksn vertex is,
 because of resolution effects, between the $\phi$ and the \kln vertex. 
%
%
%
 The $t_{\rm S}$ distribution is fit with an exponential function taking into account resolution effects
 by constructing a smearing matrix from MC.
The accuracy of the MC description is evaluated by 
increasing or reducing
the resolution by a 
scale factor $k$
according to the relation:
\begin{equation}\label{eq:resvar}
\frac{ \sigma^{\prime}_{\rm MC} }{ \sigma_{\rm MC} }= k~,
\end{equation}
where $\sigma_{\rm MC} = (t_{\rm S}^{\rm MC\, true} - t_{\rm S}^{\rm MC\, reco} )$ and $\sigma_{\rm MC}^{\prime} = (t_{\rm S}^{\rm MC \, true} - t_{\rm S}^{\prime \, \rm MC\, reco})$,
with $t_{\rm S}^{\rm MC\, reco}$ and $t_{\rm S}^{\rm MC\, true}$ the MC reconstructed and true $t_{\rm S}$, respectively.
The new reconstructed time $t_{\rm S}^{\prime \, \rm MC\, reco}$ evaluated from equation ($\ref{eq:resvar}$) is then used to construct a modified smearing matrix. 
By fixing the normalization of the distribution to data and $\tau_{\rm S}$ to its known value~\cite{kloelifetime}, 
$\tau_{\rm S}=0.89562\times 10^{-10}\, \mathrm{s}$,
a $\chi^2$ scan as a function of $k$ 
is 
performed to find the best value of the scale factor. The result is
$k = 0.9962 \pm 0.0044$,
which is compatible with unity with an uncertainty of 0.4\%.
As a cross-check, a fit with $\tau_{\rm S}$ as a free parameter (and $k=1$) yields as result 
$\tau_{\rm S}^{\mathrm{ (fit)}}=(0.8952\pm0.0030)\times 10^{-10}\, \mathrm{s}$.
\par
Similarly, 
the 
systematic uncertainty related to the $\Delta t$ resolution 
is
evaluated 
by 
increasing or reducing
%
the
resolution and correspondingly modifying the smearing matrix 
$S_{i,j}$
using the relation:
%
\begin{equation}
\frac{ \Delta t^{\rm MC \, true}   -   \Delta t^{\prime \, \rm MC\, reco} }{ \Delta t^{\rm MC \, true}   -   \Delta t^{ \rm MC\, reco}  } = 
1+\delta k~,
    \label{eq:dtk}
\end{equation}
with
$\Delta t^{\rm MC\, reco}$ and $\Delta t^{\rm MC\, true}$ the MC reconstructed and true $\Delta t$, respectively,
and $\delta k=\pm 0.75\%, \pm 1.5\%$.
The 
variation $\delta k=\pm 0.75\%$,
larger by 
almost a factor two
than 
the uncertainty on the scale factor $k$ discussed above, 
is
considered for the evaluation of
the systematic uncertainty.
%
\item
{\bf Input Physical Constants} --
In order to evaluate the 
effect induced by the uncertainty on the values of the physical constants 
($\tau_{\rm S}$, $\tau_{\rm L}$, $\Delta m$, and $\eta_{+-}$)
used in input to the fit theoretical function $I(t_1, t_2|{\bf q})$, 
different sets of physical constants 
are
randomly generated according to their uncertainty~\cite{pdg}, and the fit repeated for 
each set
of 
 constants. 
The standard deviation of the whole set of results 
is
taken as systematic uncertainty.
%
%
\end{itemize}
\par
The results on the various systematic uncertainty contributions
are summarized in table~\ref{tab:syst}. 
All 
contributions
are evaluated, except for the one on the input physical constants, as 
the semi-distance between the maximum and minimum values 
among the 
results 
obtained
with 
positive, negative, and
no variation of the considered cut or effect.
The final value of the systematic uncertainty is obtained by the sum in quadrature of the different contributions.

\begin{table}[hbt!]
    \centering
    \begin{tabular}{|c|c|c|c|c|c|c|c|}
    \hline
        &$\delta\zeta_{\rm SL}$ & $\delta\zeta_{0\bar{0}}$&$ \delta\gamma$ &$\delta\Re\omega$&$\delta\Im\omega$ &$\delta|\omega|$ & $\delta\phi_{\omega}$\\
        &$\cdot 10^2$ & $\cdot 10^{7}$&$ \cdot 10^{21}$ GeV &$\cdot 10^4$&$\cdot 10^4$ &$\cdot 10^{4}$ & $({\rm rad})$\\
    \hline
    Cut stability  & $ 0.56$ & $ 2.9$ & $ 0.33$ &$ 0.53$ & $ 0.65$ & $ 0.78$ & $0.07$\\
    $4\pi$ background & $ 0.37$ & $ 1.9$ & $ 0.22$ & $ 0.32$ & $ 0.19$& $ 0.32$ & $0.04$\\
    Regeneration & $ 0.17$ & $ 0.9$ & $ 0.10$ & $ 0.06$ & $ 0.63$& $ 0.58$ & $0.05$\\
    $\Delta t$ resolution & $ 0.18$ & $ 0.9$ & $ 0.10$ & $ 0.15$ & $ 0.09$ & $ 0.15$ & $0.02$\\
    Input phys. const.& $ 0.04$ & $ 0.2$ & $ 0.02$ & $ 0.03$ & $ 0.09$ & $ 0.07$ & $0.01$\\
    \hline
    Total & $ 0.71$ & $ 3.7$ & $ 0.42$ & $ 0.64$ & $ 0.93$& $ 1.04$ & $0.10$\\
    \hline
    \end{tabular}
    \caption{Systematic uncertainties on all decoherence {and \CPT-violating parameters.}} 
    \label{tab:syst}
\end{table}

\section{Results}
\label{sec:results}
\par
The final results for all models 
mentioned
 in section~\ref{sec:intro} are reported in the following.
\\
For the decoherence parameters $\zeta_{\rm SL}$, $\zeta_{0\bar{0}}$, and $\gamma$ they
are:
\begin{align*}
     &\zeta_{\rm SL} = (0.1 \pm 1.6_{\rm stat} \pm 0.7_{\rm syst})\cdot 10^{-2} \hbox{~with~} \chi^2/{\hbox{dof}}=11.2/10~,\\
     &\zeta_{0\bar{0}} = (-0.05 \pm 0.80_{\rm stat} \pm 0.37_{\rm syst}) \cdot 10^{-6} \hbox{~with~} \chi^2/{\hbox{dof}}=11.2/10~,\\
     &\gamma = (0.13 \pm 0.94_{\rm stat} \pm 0.42_{\rm syst} )\cdot 10^{-21} \text{~GeV}  \hbox{~with~} \chi^2/{\hbox{dof}}=11.2/10~.
\end{align*}
\par
The high precision of the
 $\zeta_{0\bar{0}}$ result with respect to $\zeta_{\rm SL}$
 can be intuitively explained
by considering 
that the overall decay,
in which both kaons decay into $\pi^+\pi^-$, is 
suppressed by \CP violation.
In quantum mechanics this conclusion is independent on the basis used in the calculation of
the decay intensity (\ref{eq:intensity}),
while
in case of a decoherence mechanism it depends on the basis
into which the initial state tends to 
factorize. 
The decay 
$\ksn \kln \rightarrow \pi^+\pi^-\pi^+\pi^-$ 
is still suppressed by \CP violation, while
the decay 
$\kn \knb \propto( \kln\ksn - \ksn\kln + \ksn\ksn-\kln\kln)\rightarrow \pi^+\pi^-\pi^+\pi^-$ has a contribution 
from $\ksn \ksn \rightarrow \pi^+\pi^-\pi^+\pi^-$ that it is not \CP suppressed, and
is copious in the region at $\Delta t \approx0$. Consequently a larger sensitivity on 
the $\zeta_{0\bar{0}}$ parameter is achieved.
\par
The $\lambda$ parameter derived from $\zeta_{\rm SL}$ \cite{bertlmann2} is:
\begin{align*}
\lambda = (0.1 \pm 1.2_{\rm stat} \pm 0.5_{\rm syst} )\cdot 10^{-16} \text{~GeV}~.
\end{align*}
As these parameters are constrained to be positive, the results can be translated into 
{$90\%$ confidence level (C.L.) }
upper limits \cite{feldman}:
\begin{align*}
    &\zeta_{\rm SL} < 0.030 ~, \\
    &\zeta_{0\bar{0}} < 1.4\cdot 10^{-6} ~, \\ 
    &\gamma < 1.8 \cdot 10^{-21}\text{ GeV } ~,\\ 
    &\lambda < 2.2 \cdot 10^{-16}\text{ GeV } ~.
\end{align*}
%
%
\par
\begin{figure}[hbt]
\centering
    \includegraphics[scale = 0.5]{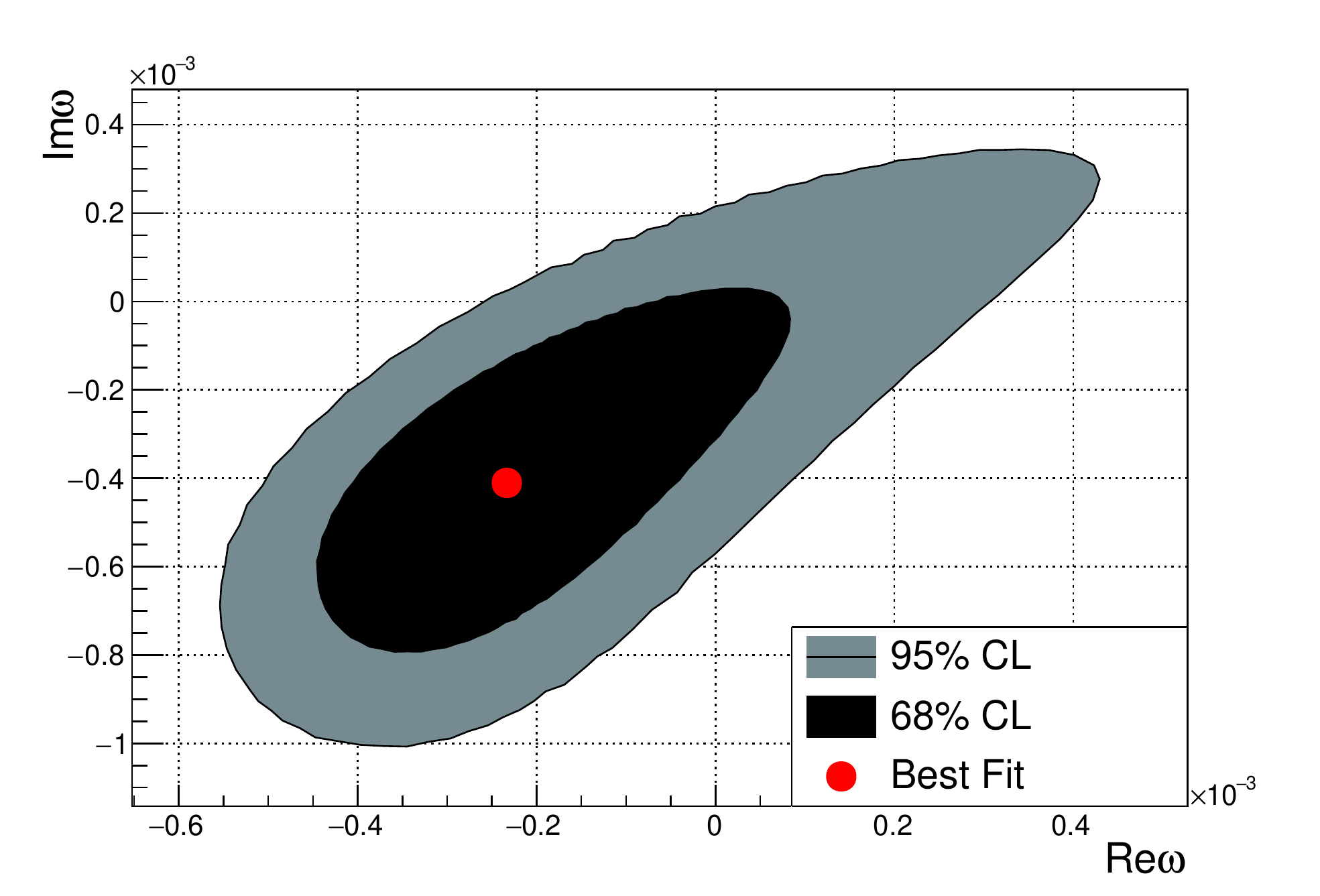}
    \caption{Contour plot of $\Im\omega$ vs $\Re\omega$ for $68\%$ and $95\%$ confidence levels.}
    \label{fig:Contour}
\end{figure}
\begin{figure}[hbt]
\centering
    \includegraphics[scale = 0.5]{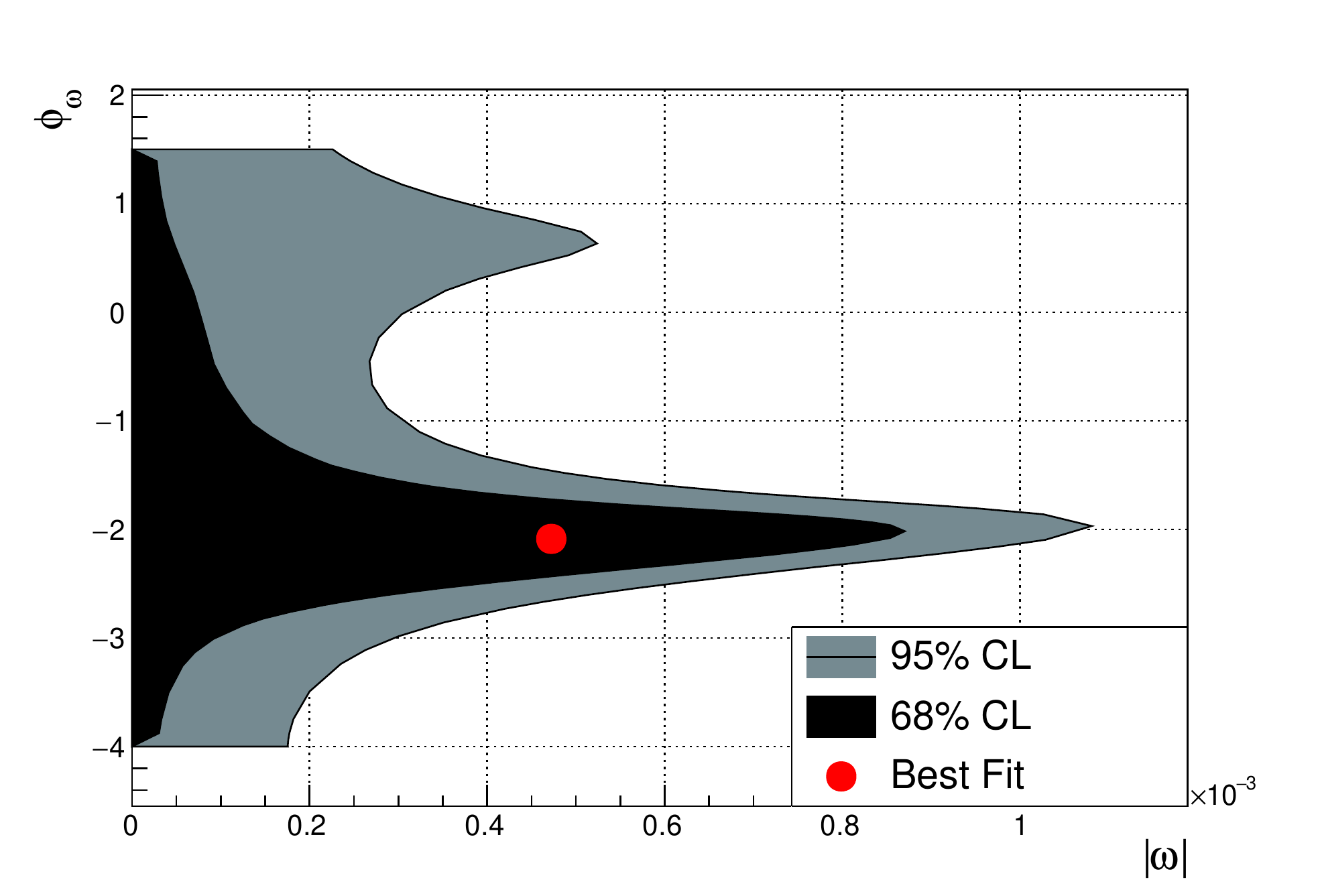}
    \caption{Contour plot of $\phi_{\omega}$ vs $|\omega|$ for $68\%$ and $95\%$ confidence levels.}
    \label{fig:Contourmodphi}
\end{figure}

The results on the complex $\omega$ parameter have been obtained by 
performing the fit in Cartesian $\{\Re\omega, \Im\omega \}$ coordinates:
\begin{align*}
     &\Re\omega = (-2.3^{+1.9}_{-1.5 {\rm stat}} \pm 0.6_{\rm syst}) \cdot 10^{-4}~,\\
     &\Im\omega = (-4.1^{+2.8}_{-2.6 {\rm stat}} \pm 0.9_{\rm syst}) \cdot 10^{-4}~, 
\end{align*}
and 
in polar $\{|\omega|, \phi_{\omega}\}$ coordinates:
\begin{align*}
     &|\omega| = (4.7 \pm 2.9_{\rm stat} \pm 1.0_{\rm syst})\cdot 10^{-4}, \\
     &\phi_{\omega} = -2.1 \pm 0.2_{\rm stat} \pm 0.1_{\rm syst}~\text{(rad)}~\hbox{~with~} \chi^2/{\hbox{dof}}=9.2/9~.
\end{align*}
The correlation coefficient between $\Re\omega$ and  $\Im\omega$ is 68\%.
The contour plot of $\Im\omega$ vs $\Re\omega$
 for the $68\%$ and $95\%$ confidence levels is shown in figure \ref{fig:Contour}.
The correlation coefficient between $|\omega|$ and  $\phi_{\omega}$ is 14\%.
The contour plot of $|\omega|$ vs $\phi_{\omega}$ 
 for the $68\%$ and $95\%$ confidence levels
is shown in figure \ref{fig:Contourmodphi}.
\par
These results 
represent a sizeable improvement with respect to previous measurements \cite{cplearzeta,cplearabc,bertlmann1},
in particular by more than a factor two with respect to previous KLOE results \cite{kloeqm2006},
and constitute the most stringent existing limits 
 on the corresponding observable effects. 
 All results are consistent with no decoherence  and no \CPT violation,
while the precision in the cases of $\gamma$ and $\omega$ parameters is reaching, or surpassing, the level of the interesting Planck's scale 
region\footnote{As mentioned in section \ref{sec:intro}, this corresponds to
$\sim 2 \times 10^{-20} \,\mbox{GeV}$ for $\gamma$, and $\sim 10^{-3}$ for $|\omega|$~.
},
the scale at which, heuristically, 
these effects may appear in the most optimistic scenarios for quantum gravity~\cite{ellis1,ellis2,ellisstring92,ellis2000,mavro1,mavro2,mavro3}.
\par
Finally from the result on $|\omega|$, using relation (\ref{eq:brksks}) and the measured value of ${{\rm BR}(\phi\rightarrow \ksn \kln)}$ \cite{pdg},
an upper limit for the branching ratio of the $\phi \rightarrow \ksn \ksn , \kln\kln$ decay can be 
derived {at 90\% C.L.}:
\begin{equation*}
    {\rm BR}(\phi\rightarrow \ksn \ksn , \kln\kln) < 2.4 \cdot 10^{-7}~.
\end{equation*}

\appendix
\section{\boldmath 
Decay intensity expressions in decoherence models}
\label{appendix}
{
\subsection*{\boldmath $\zeta_{0\bar{0}}$ model}
The decay intensity $I(t_1,t_2;\zeta_{0\bar{0}})$ is defined by introducing a factor 
$(1-\zeta_{0\bar{0}})$
multiplying the interference term of the quantum mechanical decay intensity expression in the 
$\{\kn,\knb\}$
basis:
\begin{eqnarray}
\label{eq:intensitydec00}
I(t_1,t_2;\zeta_{0\bar{0}})
&\propto&
\Big| \langle \pi^+\pi^- |T| \kn(t_1)  \rangle \langle \pi^+\pi^- |T| \knb(t_2)  \rangle \Big|^2
+\Big|\langle \pi^+\pi^- |T| \knb(t_1)  \rangle \langle \pi^+\pi^- |T| \kn(t_2)  \rangle\Big|^2
\nonumber\\
& & -2 (1-\zeta_{0\bar{0}}) 
\, \Re \Big[ \langle \pi^+\pi^- |T| \kn(t_1)  \rangle \langle \pi^+\pi^- |T| \knb(t_2)  \rangle  
\nonumber\\
& & \times 
\left( \langle \pi^+\pi^- |T| \knb(t_1)  \rangle \langle \pi^+\pi^- |T| \kn(t_2)  \rangle \right)^{\ast} 
\Big]
~.
\end{eqnarray} 
The explicit expression for $I(t_1,t_2;\zeta_{0\bar{0}})$ to lowest order in $|\eta_{+-}|$ is:
\begin{eqnarray}
\label{eq:intensitydec00}
I(t_1,t_2;\zeta_{0\bar{0}})
&\propto&
\left[ 
\left( 1- \frac{\zeta_{0\bar{0}}}{2} \right)\left( e^{-\Gamma_{\rm L} t_1 -\Gamma_{\rm S} t_2}
+ e^{-\Gamma_{\rm S} t_1 -\Gamma_{\rm L} t_2} 
-2 \, e^{ - \frac{(\Gamma_{\rm S}+\Gamma_{\rm L})}{2} (t_1+t_2)}\cos[ \Delta m (t_1-t_2)] \right) 
\right.
\nonumber\\
& & \left. +\frac{\zeta_{0\bar{0}}}{2} \left( \frac{ e^{ - \Gamma_{\rm S}(t_1+t_2)} }{|\eta_{+-}|^2}
- 2\, e^{ - \frac{(\Gamma_{\rm S}+\Gamma_{\rm L})}{2} (t_1+t_2)}\cos[ \Delta m (t_1+t_2)] 
\right)
\right]~.
%
\end{eqnarray}  
}
{
\subsection*{\boldmath $\lambda$ model}
In the model described in ref.~\cite{bertlmann2} decoherence 
is governed by a parameter $\lambda$ which is in direct correspondence to the 
parameter $\zeta_{\rm SL}$ introduced in eq.(\ref{eq:intensitydec}). The only difference in this case is that
$\zeta_{\rm SL}$
becomes time dependent (see eq.(3.3) in Ref.~\cite{bertlmann2}):
%
%
%
\begin{eqnarray}
\zeta_{\rm SL}(t_1,t_2)
= 1- e^{-\lambda \min(t_1,t_2)}~.
\label{declam}
\end{eqnarray}
Substituting expression (\ref{declam}) in eq.(\ref{eq:intensitydec}) one obtains
the decay intensity $I(t_1,t_2;\lambda)$:
\begin{eqnarray}
\label{eq:intensitydecapp}
I(t_1,t_2;\lambda) &\propto&
e^{-\Gamma_{\rm L} t_1 -\Gamma_{\rm S} t_2}
+ e^{-\Gamma_{\rm S} t_1 -\Gamma_{\rm L} t_2} 
\nonumber \\
& & -2 \left(e^{-\lambda \min(t_1,t_2)}\right) e^{ - \frac{(\Gamma_{\rm S}+\Gamma_{\rm L})}{2} (t_1+t_2)}\cos[ \Delta m (t_1-t_2)]
~.
\end{eqnarray}  
}
{
As in our case the decoherence parameter $\zeta_{\rm SL}$ in eq.(\ref{eq:intensitydec}) is measured fitting the observed $\Delta t$
distribution, comparing the decay intensities (\ref{eq:intensitydec}) 
and (\ref{eq:intensitydecapp}) after integration on the sum $(t_1+t_2)$, 
%
the following relationship holds:
\begin{eqnarray}
\zeta_{\rm SL}
=\frac{\lambda}{\Gamma_{\rm S}+\Gamma_{\rm L}+\lambda}\simeq \frac{\lambda}{\Gamma_{\rm S}}~,
\label{declam2}
\end{eqnarray}
used to evaluate $\lambda$ from the measured $\zeta_{\rm SL}$ parameter.
}
%
{
\subsection*{\boldmath $\gamma$ model}
The explicit expression for $I(t_1,t_2;\gamma)$ 
is (see Ref.~\cite{peskin}):
\begin{eqnarray}
\label{eq:intensitydecgamma}
I(t_1,t_2;\gamma)
&\propto&
\left[ 
\left( 1+ \frac{\gamma}{\Delta\Gamma |\eta_{+-}|^2} \right)\left( e^{-\Gamma_{\rm L} t_1 -\Gamma_{\rm S} t_2}
+ e^{-\Gamma_{\rm S} t_1 -\Gamma_{\rm L} t_2}  \right) 
\right.
\nonumber\\
& & \left. 
-2 \, e^{ - \frac{(\Gamma_{\rm S}+\Gamma_{\rm L})}{2} (t_1+t_2)}\cos[ \Delta m (t_1-t_2)] 
-2 \frac{\gamma}{\Delta\Gamma |\eta_{+-}|^2} e^{ - \Gamma_{\rm S}(t_1+t_2)} 
\right]~.
%
\end{eqnarray}  
}
{
\subsection*{\boldmath $\omega$ model}
The explicit expression for $I(t_1,t_2;\omega)$ 
is (see Ref.~\cite{mavro2}):
\begin{eqnarray}
\label{eq:intensitydecomega}
I(t_1,t_2;\omega)
&\propto&
\left[ 
e^{-\Gamma_{\rm L} t_1 -\Gamma_{\rm S} t_2}
+ e^{-\Gamma_{\rm S} t_1 -\Gamma_{\rm L} t_2} 
-2 \, e^{ - \frac{(\Gamma_{\rm S}+\Gamma_{\rm L})}{2} (t_1+t_2)}\cos[ \Delta m (t_1-t_2)] 
\right.
\nonumber\\
& &  +\frac{|\omega|^2}{|\eta_{+-}|^2}  e^{ - \Gamma_{\rm S}(t_1+t_2)}
+2 \frac{|\omega|}{|\eta_{+-}|} \left( e^{ - \Gamma_{\rm S} t_1} e^{ - \frac{(\Gamma_{\rm S}+\Gamma_{\rm L})}{2} t_2}
\cos[ \Delta m \, t_2 - \phi_{+-} +\phi_{\omega}]  \right.
\nonumber\\
& & \left. \left. 
- e^{ - \Gamma_{\rm S} t_2} e^{ - \frac{(\Gamma_{\rm S}+\Gamma_{\rm L})}{2} t_1}
\cos[ \Delta m \, t_1 - \phi_{+-} +\phi_{\omega}] \right)
\right]~.
%
\end{eqnarray}  
}
%
%
%

\acknowledgments
We warmly thank our former KLOE colleagues for the access to the data collected during the KLOE data taking campaign.
{We are very grateful to our colleague G.~Capon for his enlightening comments and suggestions about the manuscript.}
We thank the DA$\Phi$NE team for their efforts in maintaining low background running conditions and their collaboration during all data taking. We want to thank our technical staff: 
G.F. Fortugno and F. Sborzacchi for their dedication in ensuring efficient operation of the KLOE computing facilities; 
M. Anelli for his continuous attention to the gas system and detector safety; 
A. Balla, M. Gatta, G. Corradi and G. Papalino for electronics maintenance; 
C. Piscitelli for his help during major maintenance periods. 
This work was supported in part 
by the Polish National Science Centre through the Grants No.
2014/14/E/ST2/00262,
2014/12/S/ST2/00459,
2016/21/N/ST2/01727,
2017/26/M/ST2/00697.
%
%
%
%



\begin{thebibliography}{99}


\bibitem{epr} A.~Einstein, B.~Podolsky, N.~Rosen, {\it Can quantum mechanical description of physical reality be considered complete?},
{ \em Phys. Rev.} {\bf 47} (1935) 777.



\bibitem{schro} E.~Schr\"odinger, {\it Discussion of probability relations between separated systems}, { \em Math.
Proc.
Cambridge 
Phil.
Soc.
} {\bf 31} (1935) 555



\bibitem{leeyang} 
T.~D.~Lee, C.~N.~Yang, {\it Reported by T.~D.~Lee at the Argonne National Laboratory}, {\em unpublished} (1960),
See text and footnotes in ref.~\cite{day,inglis}. 

\bibitem{day} T.~B.~Day, {\it Demonstration of Quantum Mechanics in the large}, 
{
\em Phys. Rev.} {\bf 121} (1960) 1204.

\bibitem{inglis} D.~R.~Inglis, {\it Completeness of Quantum Mechanics and Charge-Conjugation correlations of Theta particles},
{
\em
Rev. Mod. Phys.} {\bf 33} (1961) 1.

\bibitem{lipkin68}
H.~J.~Lipkin, {\it CP violation and coherent decays of kaon pairs}, 
{\em
Phys. Rev.} {\bf 176} (1968) 1715.

\bibitem{dunietz} I.~ Dunietz, J.~Hauser, J.~L.~Rosner, {\it Proposed experiment addressing CP and CPT violation in the \kn-\knb system},
{\em Phys. Rev. D} {\bf 35} (1987) 2166.

\bibitem{buchanan} C.~D.~Buchanan, R.~Cousins, C.~Dib, R.~D.~Peccei, J.~Quackenbush,  {\it
Testing CP and CPT violation in the neutral kaon system at a $\phi$-factory}, {Phys. Rev. D} {\bf 45}  (1992) 4088.


\bibitem{isidori} G.~D'Ambrosio, G.~Isidori, A.~Pugliese, {\it CP and CPT measuremente at DA$\Phi$NE}, {\it The second DA$\Phi$NE handbook, Vol.I}
eds. L.~Maiani, G.~Pancheri, N.~Paver,  INFN-LNF, Frascati (1995) p.63.


\bibitem{didohand} A.~Di~Domenico (ed.), {\it Handbook on Neutral Kaon Interferometry at a $\phi$-factory}, {\em Frascati Phys. Ser.} {\bf 43}
INFN-LNF, Frascati, (2007); available online at http://www.lnf.infn.it/sis/frascatiseries/Volume43/volume43.pdf~.


\bibitem{ktviol} J.~Bernabeu, A.~Di~Domenico, P.~Villanueva-Perez,
{\it Direct test of time reversal symmetry in the entangled neutral kaon
system at a $\phi$-factory}, 
{\em
Nucl. Phys. B} {\bf 868} (2013) 102.

\bibitem{kcptviol} J.~Bernabeu,  A.~Di~Domenico, P.~Villanueva-Perez,
{\it Probing CPT in transitions with entangled neutral kaons},
{\em
JHEP} {\bf 10} (2015) 139.

\bibitem{bertlmann1} R.~A.~Bertlmann, W.~Grimus, B.~C.~Hiesmayr, 
{\it Quantum mechanics, Furry's hypothesis, and a measure of decoherence in the \kn\knb system},
{\em Phys. Rev. D} {\bf 60} (1999) 114032.

\bibitem{furry} W.~H.~Furry, {\it Note on the Quantum-Mechanical theory of measurement}, {\em Phys.~Rev.} {\bf 49} (1936) 393.

\bibitem{eberhard} P.~H.~Eberhard, Tests of Quantum Mechanics at a $\phi$-factory, 
{\it The second DA$\Phi$NE handbook, Vol.I}
eds. L.~Maiani, G.~Pancheri, N.~Paver,  INFN-LNF, Frascati (1995) p.99.
%


\bibitem{bertlmann2} R.~A.~Bertlmann, K.~Durstberger, B.~C.~Hiesmayr, 
{\it Decoherence of entangled kaons and its connection to entanglement measures},
{\em Phys. Rev. A} {\bf 68} (2003) 012111.


\bibitem{hawk2} S.~Hawking, {\it The Unpredictability of Quantum Gravity}, {\em Commun. Math. Phys.} {\bf 87} (1982)
395.
\bibitem{wald} R.~Wald, {\it Quantum Gravity and Time reversibility}, { \em Phys. Rev. D} {\bf 21} (1980) 2742.

\bibitem{ellis1} J.~Ellis, J.~S.~Hagelin, D.~V.~Nanopoulos,  M.~Srednicki, {\it Search for violations of quantum mechanics}, {\em Nucl. Phys. B} {\bf241} 
(1984) 381.

\bibitem{ellis2} J.~Ellis, J.~L.~Lopez,  N.~.E.~Mavromatos, D.~V.~Nanopoulos, 
{\it Precision tests of CPT symmetry and quantum mechanics in the neutral kaon system}, {\em Phys. Rev. D} {\bf 53} (1996)
3846.

\bibitem{ellisstring92} 
J.~Ellis, N.~E.~Mavromatos, D.~V.~Nanopoulos,  
{\it String theory modifies quantum mechanics}, {\it Phys. Lett. B} {\bf 292} (1992) 
37.
%
\bibitem{ellis2000} J.~Ellis, N.~E.~Mavromatos, D.~V.~Nanopoulos,  
{\it How large are dissipative effects in noncritical Liouville string theory?}, {\it Phys. Rev. D} {\bf 63} (2000) 
024024.


\bibitem{peskin} P.~Huet, M.~E.~Peskin,  {\it Violation of CPT and quantum mechanics in the \knknb system}, {\em Nucl. Phys. B} {\bf434}
(1995) 3.



\bibitem{benattiall} F.~Benatti, F.~Floreanini, {\it Completely positive dynamical maps and the neutral kaon system},  
{\em Nucl. Phys. B} {\bf 488} (1997) 335.


\bibitem{mavro1} J.~Bernabeu, N.~Mavromatos, J.~Papavassiliou,
{\it Novel Type of CPT Violation for Correlated Einstein-Podolsky-Rosen States of Neutral Mesons}, {\em Phys. Rev. Lett.}
{\bf 92} (2004) 131601.

\bibitem{mavro2} J.~Bernabeu, N.~E.~Mavromatos, J.~Papavassiliou, A.~Waldron-Lauda, 
{\it Intrinsic CPT violation and decoherence for entangled neutral mesons}, {\em Nucl. Phys. B} {\bf 744} (2006) 180.

\bibitem{mavro3} J.~Bernabeu, N.~E.~Mavromatos, S.~Sarkar,
{\it Decoherence induced CPT violation and entangled neutral mesons}, {\it Phys. Rev. D} {\bf 74} (2006) 045014.


\bibitem{pdg} P.~Zyla {et al.}, (Particle Data Group), {\it Review of Particle Physics}, {\em Prog. Theor. Exp. Phys.} {\bf 2020} (2020) 083C01.


\bibitem{kloekkg} F.~Ambrosino et al., {\it Search for the decay $\phi\rightarrow\kn\knb\gamma$ with the KLOE experiment},
{\it Phys. Lett. B} {\bf 679} (2009) 10. 



\bibitem{kloeqm2006} F.~Ambrosino {et al.},
{\it First observation of quantum interference in the process $\phi\rightarrow\ksn\kln\rightarrow \pi^+\pi^-\pi^+\pi^-$: a Test of quantum mechanics and CPT symmetry},
{\em
Phys. Lett. B} {\bf 642} (2006) 315.

\bibitem{kloecpt2013} 
D.~Babusci {et al.}, {\it Test of CPT and Lorentz symmetry in entangled neutral kaons with the KLOE experiment},
{\em Phys. Lett. B} {\bf 730} (2014) 89.



%
%
%
%
%
%
%
\bibitem{dafne1} A.~Gallo {et al.}, {\it DA$\Phi$NE status report},
{\em
Conf. Proc.} {\bf C060626} (2006) 604.
%
{
\bibitem{dafne2} 
~M.~Zobov {et al.}, {\it Test of crab-waist collisions at the DA$\Phi$NE $\Phi$ factory},
{\em
Phys. Rev. Lett.} {\bf 104} (2010) 174801.
%
\bibitem{dafne3} 
~C.~Milardi { et al.}, {\it High luminosity interaction region design for collisions inside high field detector solenoid},
{\em
JINST} {\bf 7} (2012) T03002.
}

%
%
%
%
%
%

%
%
%

\bibitem{dc} M.~Adinolfi et al., {\it The tracking detector of the KLOE experiment}, Nucl. Instr. and Meth. A {\bf 488} (2002) 51.
\bibitem{emc}M.~Adinolfi et al., {\it The KLOE electromagnetic calorimeter}, Nucl. Instr. and Meth. A {\bf 482} (2002) 364.
\bibitem{trigger} M.~Adinolfi et al., {\it The trigger system of the KLOE experiment}, Nucl. Instr. and Meth. A {\bf 492} (2002) 134.
\bibitem{offline} F.~Ambrosino et al., {\it Data handling, reconstruction and simulation for the KLOE experiment}
Nucl. Instr. and Meth. A {\bf 534} (2004) 403.




%
%



\bibitem{qmadd} A.~Di~Domenico, {\it Testing quantum mechanics in the neutral kaon system at a $\phi$-factory}, {\em Nucl. Phys. B} {\bf 450} (1995) 293.
\bibitem{baldini} R.~Baldini, A.~Michetti, {\it \kln interactions and \ksn regeneration in KLOE}, LNF-96-008 IR (1996)
(available online at http://cds.cern.ch/record/316977/files/SCAN-9701034.pdf).
%


\bibitem{iza} I. Balwierz,
{\it Measurement of the neutral kaon regeneration cross-section in beryllium at P=110 MeV/c with the KLOE detector},
diploma thesis, Jagiellonian university, Krakow, (2011),
(available online at http://koza.if.uj.edu.pl/staff/ibalwierz)

\bibitem{kloelifetime}
F.~Ambrosino et al., {\it Precision measurement of the $\ksn$ meson lifetime with the KLOE detector},
{
\it 
Eur. Phys. J. C}  {\bf 71} (2011) 1604.

\bibitem{feldman} G.~J.~Feldman, R.~D.~Cousins, {\it Unified approach to the classical statistical analysis of small signals},
{\em Phys. Rev. D} {\bf 57} (1998) 3873.


\bibitem{cplearzeta} 
A.~Apostolakis et al. {\it An EPR experiment testing the nonseparability of the $K^0$ anti-$K^0$ wave function}, {\em Phys.Lett. B} 
{\bf 422} (1998) 339.

\bibitem{cplearabc} R.~Adler {et al.}, {\it Tests of CPT symmetry and quantum mechanics with experimental data
from CPLEAR}, {\em Phys. Lett. B} {\bf 364} (1995) 239.

%
%
%
%
%
%
%
%
%
%





%
%
%
%




\end{thebibliography}
\end{document}